\documentclass[aps, reprint, twocolumn, showpacs,superscriptaddress]{revtex4-2}
\usepackage{amsmath}
\usepackage[
colorlinks=true,
pdfborder={0 0 0},
breaklinks,
allcolors=blue
]{hyperref}
\usepackage{graphicx}
\usepackage{xcolor}
\usepackage{newtxtext}
\usepackage{orcidlink}
\usepackage{upgreek}
\usepackage{amsmath}

\newcommand{\vect}{\boldsymbol}
\newcommand{\diff}{\mathrm{d}}
\newcommand{\Eqref}[1]{Eq.~\eqref{#1}}

\newcommand{\figref}[1]{Fig.~\ref{#1}}
\newcommand{\Figref}[1]{Fig.~\ref{#1}}
\newcommand{\kbT}{k_\mathrm{B}T}

\newcommand{\lyapunov}{\mathcal{L}}

\newcommand{\mob}{\Lambda}

\newcommand{\phiAbar}{\bar{\phi}_{A}}
\newcommand{\phiBbar}{\bar{\phi}_{B}}
\newcommand{\phiCbar}{\bar{\phi}_{C}}
\newcommand{\phiDbar}{\bar{\phi}_{D}}
\newcommand{\phiEbar}{\bar{\phi}_{E}}
\newcommand{\phiFbar}{\bar{\phi}_{F}}

\newcommand{\br}{\vect{r}}

\newcommand{\phibar}{\bar{\phi}}

\newcommand{\Lbeta}{{L_\beta}}

\newcommand{\Nc}{{N}}
\newcommand{\Nr}{{S}}

\newcommand{\Vbeta}{{V_\beta}}

\newcommand{\sectionCustom}[1]{}
\newcommand{\subsectionCustom}[1]{}

\DeclareRobustCommand{\react}{\ensuremath{\rightleftharpoons}}

\begin{document}
\title{Coexistence of patterned phases in chemically active multicomponent mixtures}

\author{Chengjie Luo\,\orcidlink{0000-0001-8443-0742}}
\affiliation{Max Planck Institute for Dynamics and Self-Organization, Am Faßberg 17, 37077 Göttingen, Germany}

\author{Yicheng Qiang\,\orcidlink{0000-0003-2053-079X}}
\affiliation{Max Planck Institute for Dynamics and Self-Organization, Am Faßberg 17, 37077 Göttingen, Germany}

\author{Guido L. A. Kusters\,\orcidlink{0000-0002-6004-9768}}
\affiliation{Max Planck Institute for Dynamics and Self-Organization, Am Faßberg 17, 37077 Göttingen, Germany}

\author{David Zwicker\,\orcidlink{0000-0002-3909-3334}}
\thanks{Contact author:  \href{mailto:david.zwicker@ds.mpg.de}{david.zwicker@ds.mpg.de}}
\affiliation{Max Planck Institute for Dynamics and Self-Organization, Am Faßberg 17, 37077 Göttingen, Germany}

\begin{abstract}
Chemically active mixtures exhibit complex patterns that emerge from the interplay of physical interactions and reactions among components. Individually, these two processes are well-understood: Physical interactions can give rise to phase separation, whereas reactions can form reaction-diffusion patterns. To understand the combination of both processes, we identify a Lyapunov functional for a class of chemical reactions. By minimizing this functional, we identify a generalized Gibbs phase rule that governs the number of coexisting patterns, and we demonstrate that complex patterns can be created by the modular combination of independent phases. Our theory unveils complex stationary patterns in chemically active mixtures and provides a framework for analyzing more complex systems.
\end{abstract}
\maketitle

\sectionCustom{Introduction}

Chemically active mixtures are ubiquitous, ranging from the organization of living cells~\cite{weber2019physics} to complex reaction-diffusion patterns in synthetic applications~\cite{cross1993pattern}.
Since such systems are driven away from equilibrium, final states are often difficult to predict.
In contrast, passive systems are comparatively simple: They relax to equilibrium where different macroscopic phases can coexist.
For a system with $N$ components and $C$ constraints, such as electroneutrality, Gibbs phase rule implies that there are at most $N-C$ such phases~\cite{Gibbs1876}.
This result provides powerful predictions for passive multicomponent mixtures~\cite{mao2020designing, qiang2024scaling}, but similarly strong results are absent for chemically active mixtures. 

Even in simple cases, chemically active mixtures can exhibit complex behavior.
For example, systems with just two components exhibit stationary states that are either homogeneous or form a regular pattern whose length scale is controlled by the reactions~\cite{glotzer1994monte,Christensen1996,Zwicker2015,zwicker2022intertwined}.
Adding one component, such a patterned region can also coexist with a homogeneous region~\cite{Bauermann2024}.
These examples can be understood qualitatively by interpreting the effect of reactions as long-range repulsion~\cite{Christensen1996, Liu1989}, which is known to yield regular patterns~\cite{Ohta1986,Muratov2002,thewes2025phase}, but it is unclear whether these ideas generalize to multicomponent mixtures.
Such mixtures have been studied using linear stability analysis~\cite{turing1952chemical,cross1993pattern, Carati1997, Aslyamov2023, Avanzini2024}, but this method generally cannot predict stationary states.

\begin{figure}[t]
    \centering
    \includegraphics[width=\linewidth]{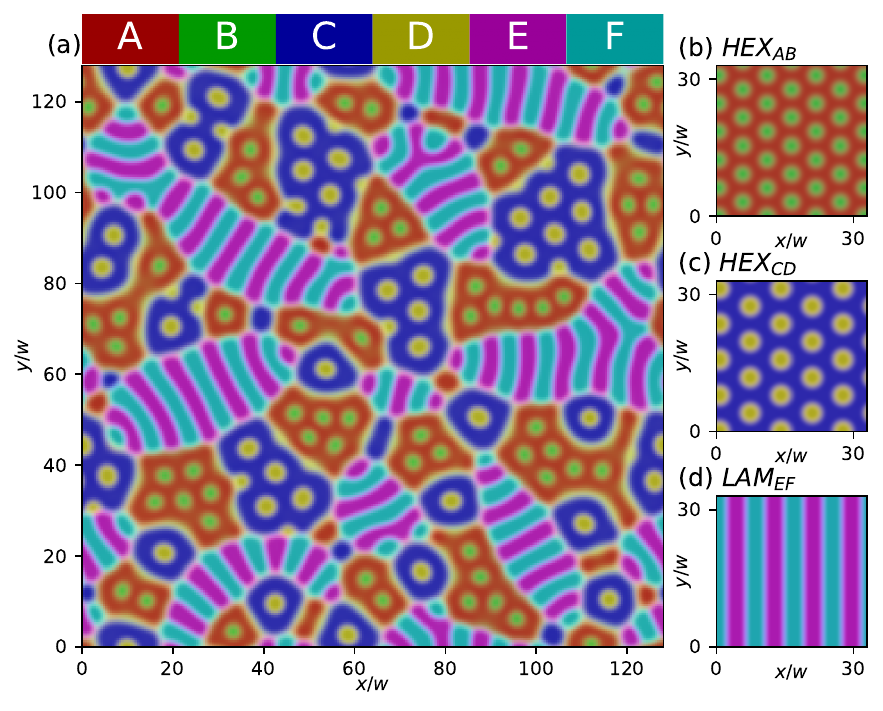}
    \caption{%
    Complex, multi-scale pattern in a mixture of six components $A$--$F$ with reactions $A {\react} B$, $C {\react} D$, and $E {\react} F$.
    (a)~Snapshot of volume fractions $\phi_i$ (shown by overlaying colors according to legend at the top) at $t=3390 w^2/D_0$ for a simulation of \Eqref{eqn:pde}.
    (b)--(d)~Equilibrium profiles of the three coexisting phases obtained by minimizing the Lyapunov functional $\lyapunov$ given by \Eqref{eqn:lyapunov}.
    (a--d)~Free energy given by \Eqref{eqn:free_energy} and bare diffusion matrix $M_{ij}=\delta_{ij}m_{i}D_0$, where $m_i=\{1, r_1^{-1}, 1, r_2^{-1}, 1, r_3^{-1}\}$ with $r_1=r_2=3$ and $r_3=1$.
    Reaction rates are $k_{AB}=r_1k_{BA}=0.85k_0$, $k_{CD}=r_2k_{DC}=0.38k_0$, and $k_{EF}=r_3k_{FE}=0.21k_0$, where $k_0=D_0/w^2$.
    Parameters are  $\chi = 6$, $\phiAbar{=}\phiCbar{=}0.24$, $\phiBbar{=}\phiDbar{=}0.08$, and $\phiEbar{=}\phiFbar{=}0.18$. %
    }
    \label{fig:snapshot_6comp}
\end{figure}

We here analyze multicomponent mixtures and identify which reactions can be mapped to long-range interactions.
We demonstrate that each reaction imparts a constraint on the system, and that an analogy to Gibbs phase rule constrains stationary states.
Moreover, we show how reaction rates control pattern morphologies, which ultimately allows us to explain complex patterns like that shown in \figref{fig:snapshot_6comp}.

\sectionCustom{Results}

\subsectionCustom{Mapping to an equilibrium system with long-range interactions}
Our model describes an isothermal, incompressible mixture of $N$ different components enclosed in a system of constant volume~$V$.
The state of the system is characterized by the volume fractions $\phi_i(\vect{r},t)$ for all components~$i$ at position $\vect{r}$ at time~$t$, which obey the no-void condition $1=\sum_i \phi_i$.
The thermodynamic properties of the system are encoded by the free energy functional $F[\{\phi_i\}]$, where we measure all energies in units of $\kbT$.
Equilibrium states are characterized by balanced chemical potentials $ \mu_j = \nu \delta F/\delta \phi_j$, where $\nu$ is a molecular volume, which we assume to be the same for all components for simplicity.
In passive systems with conserved average fractions $\bar\phi = V^{-1}\int\phi\diff V$, minimization of $F$ can lead up to $N$ coexisting, macroscopic, homogeneous phases.

Driven chemical reactions can affect the system's state.
In the simplest case, the dynamics of $\phi_i$ are governed by
\begin{align}
    \frac{\partial \phi_i}{\partial t} = 
        \sum_{j=1}^{N} \mob_{ij} \nabla^2 \mu_j + 
        \sum_{j=1}^{N} k_{ij} \phi_j
        \;,
        \label{eqn:pde}
\end{align}
where the two terms on the right account for spatial fluxes and reactions, respectively.
Here, spatial fluxes are driven by gradients in chemical potentials~$\mu_j$, proportional to the positive-definite mobility matrix $\mob_{ij}$~\cite{onsager,Julicher_2018}, which needs to satisfy $\mob_{ij}=M_{ij}-\sum_lM_{il}\sum_kM_{kj}/\sum_{kl}M_{kl}$ with bare mobilities $M_{ij}$, so \Eqref{eqn:pde} is consistent with the no-void condition~\cite{Zwicker2025}.
In contrast, reactions are described by a constant rate matrix $k_{ij}$, which satisfies $\sum_i k_{ij}=0$ to ensure mass conservation.

Since the reactions described by \Eqref{eqn:pde} are not derived from the free energy~$F$, they must be actively driven~\cite{kirschbaum2021controlling,Zwicker2022a} and can thus induce complex spatial patterns.
As an example, \figref{fig:snapshot_6comp} shows a snapshot of a simulation of \Eqref{eqn:pde}, which contains different types of droplets, striped regions, and regular patterns.
In essence, the system exhibits multiple length scales, which is not surprising since combining rates $k_{ij}$ with mobilities $\mob_{ij}$ leads to many reaction-diffusion length scales, $(\mob_{ij}/k_{nm})^{1/2}$, which could determine structures in the final pattern. 
This observation raises the question of what patterns are possible and how many length scales can be observed simultaneously in such chemically active mixtures.

To systematically determine which steady-state patterns are possible, we construct a Lyapunov functional $\lyapunov$ whose minima correspond to stable stationary states of \Eqref{eqn:pde}.
Such a functional exists when one can find generalized charges $q_i^{(s)}$ (with dimensions of inverse length) satisfying
\begin{align}
     -k_{ij}=\sum_{l=1}^{N} \mob_{il}\sum_{s=1}^S q_l^{(s)} q_j^{(s)}
    \label{eqn:decomposition}
    \;,
 \end{align}
 which implies that the left-hand side can be decomposed into two symmetric positive semi-definite matrices, $\mob$  and $\boldsymbol{q}\boldsymbol{q}^\top$.
 Note that the associated rank $S$ corresponds to the number of independent charge flavors representing independent reactions.
If such a decomposition exists, $\lyapunov$ takes the form (Appendix \ref{sec:dynamics})
\begin{align}
\lyapunov= F
    + \frac{1}{\nu}\!\int\! \sum_{s=1}^S \!\left[
        \psi^{(s)}\! \sum_{i=1}^N q_i^{(s)} \phi_i
        -\tfrac{1}{2} |\nabla \psi^{(s)}|^2
    \right]\!\mathrm{d}V
    \;,
    \label{eqn:lyapunov}
\end{align}
where the auxiliary fields $\psi^{(s)}$ satisfy the Poisson equations $\nabla^2 \psi^{(s)} = - \sum_{i} q_i^{(s)} \phi_i$.
In this case, gradient dynamics, $\partial_t \phi_i = \sum_j\mob_{ij} \nabla^2 (\nu\delta \lyapunov/\delta \phi_j)$, are identical to \Eqref{eqn:pde} if the average fractions~$\bar\phi_i$ obey (Appendix \ref{sec:dynamics})
\begin{align}
    \sum_{j=1}^N \left(\sum_{l=1}^N \sum_{s=1}^S \mob_{il}\, q^{(s)}_l q^{(s)}_j\right) \bar{\phi}_j =0
    \;,
    \label{eqn:constraint}
\end{align}
to capture that spatially averaged reactions vanish in stationary state, $\sum_j k_{ij}\bar\phi_j = 0$.
Consequently, we can study the minima of $\lyapunov$ to learn about stationary states of \Eqref{eqn:pde}.

\subsectionCustom{Phase coexistence in a minimal model}

The Lyapunov functional $\lyapunov$ can be interpreted as the free energy of an equilibrium system in which each charge flavor mediates a long-range, Coulomb-like interaction.
\Eqref{eqn:constraint} then provides $S$ independent constraints on the mean compositions $\bar\phi_i$, in direct analogy to electroneutrality conditions in electrostatic systems.
Since $\lyapunov$ can be interpreted as the free energy of an equilibrium system, Gibbs phase rule applies~\cite{Gibbs1876}: an incompressible $N$-component mixture at fixed temperature with $S$ additional  constraints supports at most
\begin{align}
    P_{\max} = N - S
    \label{eqn:phase_rule}
\end{align}
coexisting phases.
In general, each phase may be either homogeneous \emph{or} spatially patterned (i.e., a periodic steady state with a finite characteristic wavelength, such as lamellar or hexagonal order).
Importantly, these predictions also hold for stationary states of \Eqref{eqn:pde}%
, which we next test for progressively more complex systems.

\emph{Two components, one reaction ($N{=}2$, $S{=}1$)}---%
The simplest case is a system comprising two components $A$ and $B$ with a single reaction $A\react B$.
In this case, the Lyapunov functional~$\mathcal L$ essentially reduces to the Ohta--Kawasaki model~\cite{Ohta1986}, and \Eqref{eqn:phase_rule} predicts $P_{\max}=1$.
As an example, we choose the diffusive mobilities $M_{AA} = D_0$, $M_{BB} = D_0/r_1$, and $M_{AB}=M_{BA}=0$ with base diffusivity $D_0$, and map the reaction rates $k_{AB}$ and $k_{BA}$ onto a single charge flavor with $q_A = \sqrt{k_{BA}/D_0}$ and $q_B = -q_A r_1$ for $r_1 = k_{AB}/k_{BA}$ (see Appendix \ref{sec:mapping}).
Indeed, this system exhibits either a homogeneous state (weak repulsion, strong reactions) or a single patterned phase (strong repulsion, weak reactions), but never coexistence~\cite{Zwicker2015, qiao2026active,Artem2025electrostatically}.

\emph{Three components, one reaction ($N{=}3$, $S{=}1$)}---%
Adding a third species $C$ that does not participate in the reaction, but repels $A$ and $B$, implies $P_{\max}=2$.
In this case, a homogeneous phase enriched in $C$ can now coexist with either a homogeneous or patterned $AB$-rich phase (see \cite{Bauermann2024} and Appendix \ref{sec:addition}).
However, we could not find two distinct coexisting \emph{patterned} phases with just one reaction.

\begin{figure}[t]
    \centering
    \includegraphics[width=\linewidth]{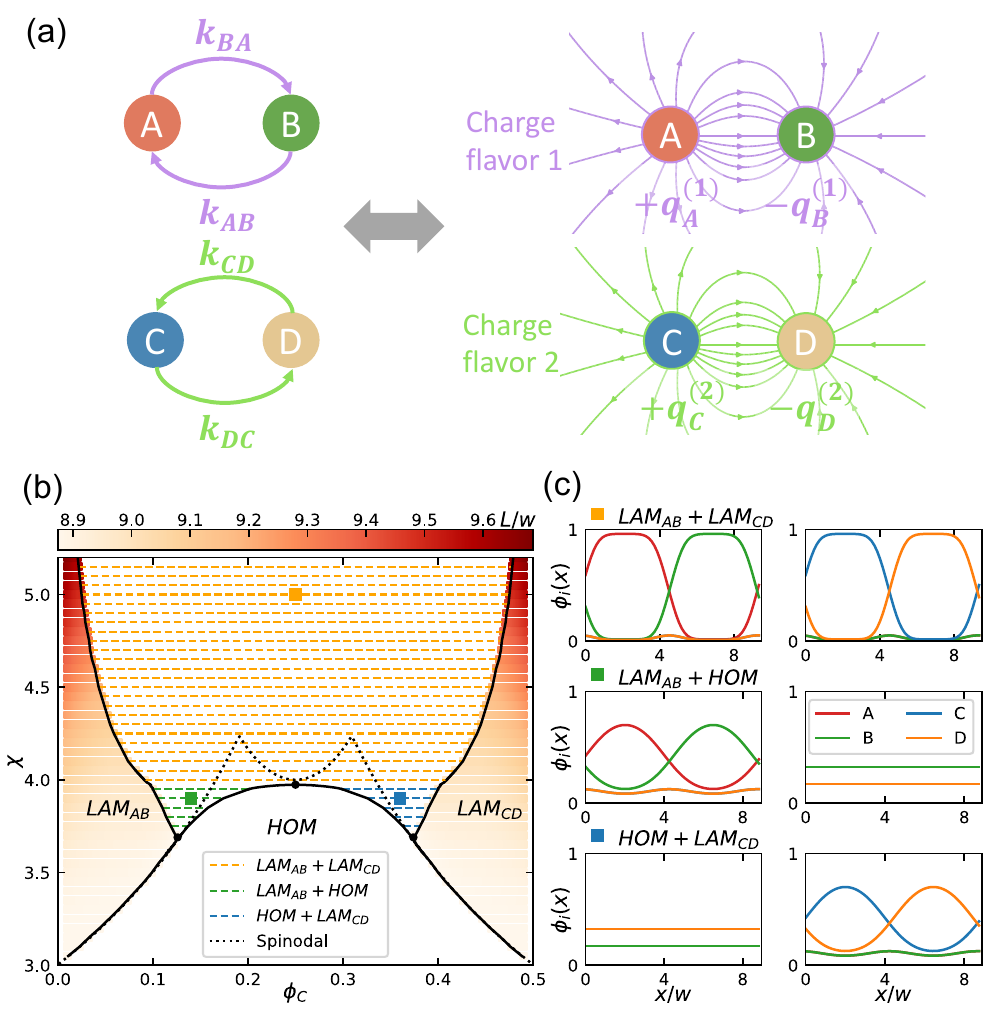}    
    \caption{
        (a) Schematic of the mapping of two linear reactions to corresponding charges mediating long-range repulsion.
        (b) State diagram from minimizing $\mathcal L$ given by \Eqref{eqn:lyapunov} as a function of the average fraction~$\phiCbar$ and interaction strength~$\chi$. 
        Length scales~$L$ of pure lamellae phases ($LAM_{AB}$ and $LAM_{CD}$) are indicated by the colorcode. 
        The homogeneous phase ($HOM$) is unstable above the spinodal (dotted black line).
        Orange dashed tie lines indicate coexistence of two lamellar phases.
        Green (blue) dashed tie lines indicate coexistence of a lamellar phase $L_{AB}$ ($L_{CD}$) and a homogeneous phase.
        (c) Exemplary volume fraction profiles for the three phase coexistence regimes.
        Parameters are indicated by colored squares in panel b (top to bottom:  $\phiCbar=0.25, \chi=5.0$; $\phiCbar=0.14, \chi=3.9$; $\phiCbar=0.36, \chi=3.9$).
        (b, c) Model parameters are $q^{(1)}_A=-q^{(1)}_B=q^{(2)}_C=-q^{(2)}_D=0.354 w^{-1}$.%
        }
    \label{fig:figure2}
\end{figure}

\emph{Four components, two reactions ($N{=}4$, $S{=}2$)}---%
To realize coexistence of two patterned phases, we introduce a second reaction pair by considering species $A$, $B$, $C$, and $D$ with reactions $A\react B$ and $C\react D$ (\figref{fig:figure2}a).
Since the two reactions do not share any components, the two charge flavors decouple:
Flavor~1 involves only $A$ and $B$, whereas flavor~2 involves only $C$ and $D$ (see \figref{fig:figure2}a and Appendix \ref{sec:mapping} for explicit expressions).
Consequently, we have $S{=}2$ and expect up to two phases ($P_{\max}=2$), which could both be patterned.

To test our predictions, we consider a concrete system described by a Flory--Huggins free energy~\cite{Flory1942,Huggins1941}
\begin{equation}
    F = \frac{1}{\nu}\int \Big[ \sum_{i} \phi_i \ln \phi_i 
    + \frac{\chi}{2}\sum_{i\neq j} \phi_i \phi_j 
    + \frac{w^2}{2} \sum_{i} |\nabla \phi_i|^2
    \Big] \diff V
    \;,
    \label{eqn:free_energy}
\end{equation}
where $i,j \in \{A,B,C,D\}$ and we enforce the no-void condition $\sum_i\phi_i=1$ using a Lagrange multiplier.
For simplicity, we consider equal repulsion $\chi>0$ between all species, and a single parameter $w$ controls the width of interfaces.
The neutrality constraints imply that each reaction must balance separately, which we capture by the associated charge asymmetry ratios $r_1={k_{AB}}/{k_{BA}} = \bar\phi_A/\bar\phi_B$ and $r_2={k_{CD}}/{k_{DC}}=\bar\phi_C/\bar\phi_D$.
These two constraints, together with the no-void condition, imply that only one average fraction can be chosen freely, where we choose $\phiCbar$ without loss of generality.

To get an intuition for the behavior of the system, we first consider symmetric reactions ($r_1=r_2=1$) and vary $\chi$.
We minimize $\mathcal L$ via a Gibbs-ensemble method (Appendix \ref{sec:Gibbs}) to obtain the phase diagram (\figref{fig:figure2}b).
As expected, the system is homogeneous ($HOM$, white region) at weak repulsion~$\chi$.
However, spatially periodic states emerge above a critical value of $\chi$: We observe a lamellar phase enriched in $A$ and $B$ ($LAM_{AB}$) when $\phiCbar$ is small, or enriched in $C$ and $D$ ($LAM_{CD}$) when $\phiCbar$ is large.
These patterned phases arise from the competition between local repulsion and effective long-range attraction mediated by the reactions~\cite{thewes2025phase}.
The transition from homogeneous to  lamellar states is continuous since the spinodal of the homogeneous state (dotted line) coincides with the phase boundary and the amplitude of the patterned phase vanishes at the boundary (\figref{fig:amplitude_chi}).
At even stronger repulsion, we observe coexistence between multiple phases.
Depending on composition, a lamellar phase coexists with the homogeneous phase ($LAM_{AB}{+}HOM$ or $HOM{+}LAM_{CD}$; \figref{fig:figure2}c, bottom two panels), or, for yet larger $\chi$, both lamellar phases coexist ($LAM_{AB}{+}LAM_{CD}$; \figref{fig:figure2}c, top panel), each with enhanced compositional contrast but nearly unchanged periodicity~\cite{luo2025theory}.
All three phases meet at a triple point ($\phiCbar^*{=}0.25$, $\chi^*{\approx}3.97$), which has measure zero in the $(\phiCbar,\chi)$ plane and is thus consistent with $P_{\max}=2$.
Taken together, these results confirm that chemically active mixtures with linear reactions exhibit at most $P_{\max} = N-S$ phases, which each may be homogeneous or patterned.

\subsectionCustom{Controlling phase morphology via reaction rates}

\begin{figure*}[t]
    \centering
    \includegraphics[width=\linewidth]{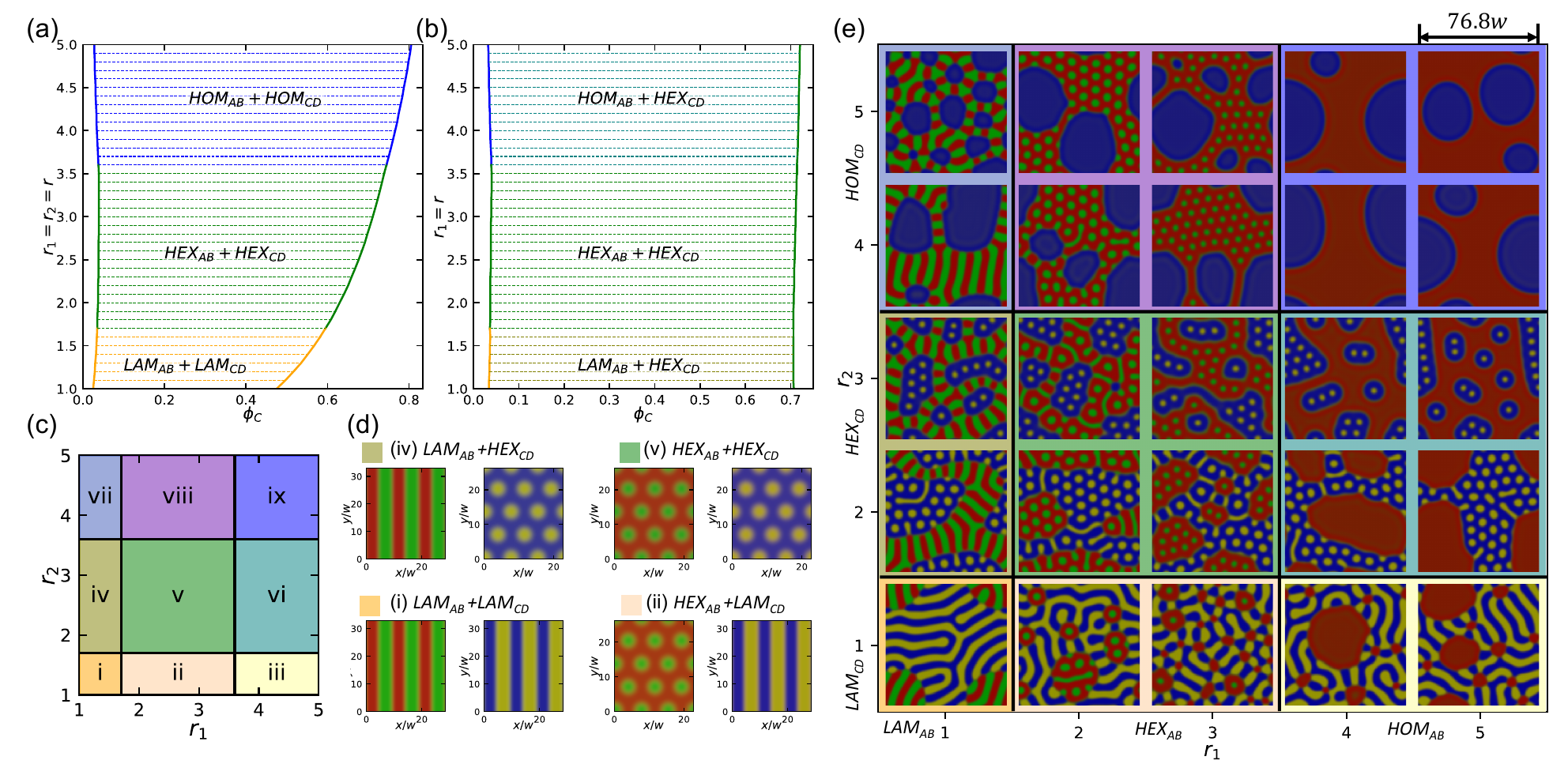}    
    \caption{%
        Reaction rate ratios~$r_i$ control phase morphology. 
        (a)~State diagram as a function of $\phiCbar$ and rate ratio $r=r_1=r_2$. %
        Increasing $r$ drives transitions from coexisting lamellae ($LAM_{AB}{+}LAM_{CD}$) through hexagonal phases ($HEX_{AB}{+}HEX_{CD}$) to homogeneous coexistence ($HOM_{AB}{+}HOM_{CD}$).
        (b)~State diagram as a function of $\phiCbar$ and $r_1$ for fixed $r_2=3$, showing that the $AB$ phase transitions independently while the $CD$ phase remains hexagonal.
        (c)~Full $(r_1, r_2)$ state diagram at $\phiCbar=0.35$.
        Roman numerals label the $3{\times}3{=}9$ distinct two-phase states:
        (i)~$LAM_{AB}{+}LAM_{CD}$,
        (ii)~$HEX_{AB}{+}LAM_{CD}$,
        (iii)~$HOM_{AB}{+}LAM_{CD}$,
        (iv)~$LAM_{AB}{+}HEX_{CD}$,
        (v)~$HEX_{AB}{+}HEX_{CD}$,
        (vi)~$HOM_{AB}{+}HEX_{CD}$,
        (vii)~$LAM_{AB}{+}HOM_{CD}$,
        (viii)~$HEX_{AB}{+}HOM_{CD}$,
        (ix)~$HOM_{AB}{+}HOM_{CD}$.
        (d)~Equilibrium profiles for the four states with two patterned phases:
        (i)~$r_1{=}r_2{=}1$;
        (ii)~$r_1{=}3$, $r_2{=}1$;
        (iv)~$r_1{=}1$, $r_2{=}3$;
        (v)~$r_1{=}r_2{=}3$.
        (e)~Snapshots of simulations of \Eqref{eqn:pde} at $t\approx10^4 w^2/D_0$ for various $(r_1, r_2)$, confirming the morphologies predicted in panel c.
        (a--e) Model parameters are $\chi=5$, $q^{(1)}_A=q^{(2)}_C=0.354 w^{-1}$, $\phiCbar=0.35$ (for panels c--e), and $M_{ij}=\delta_{ij}m_iD_0$ with $m_i=\{1, r_1^{-1}, 1, r_2^{-1}\}$ (for panel e).
           }
    \label{fig:figure3}
\end{figure*}

We next show how the \emph{morphology} of each patterned phase is controlled by the reaction rates, and particularly the ratios $r_1$ and $r_2$.
Physically, these ratios set the compositional asymmetry within each reaction pair: symmetric charges ($r=1$) thus favor lamellar order, while increasing asymmetry shifts the equilibrium toward hexagonal (droplet) phases and, ultimately, homogeneous states.

To study the influence of reactions, we fix $\chi = 5$ to have coexisting patterned phases and consider $r_1=r_2\equiv r$ (\figref{fig:figure3}a).
When increasing $r$, we encounter three different regimes: coexisting lamellae ($LAM_{AB}{+}LAM_{CD}$) for $1 < r \lesssim 1.6$, coexisting hexagonal phases ($HEX_{AB}{+}HEX_{CD}$) for $1.6 \lesssim r \lesssim 3.5$, and two homogeneous phases ($HOM_{AB}{+}HOM_{CD}$) for $r \gtrsim 3.5$.
The transition from lamellar to hexagonal phases reflects the growing charge asymmetry controlled by $r$: as the minority species is depleted, the patterned phase evolves from stripes to droplets, mirroring the lamellar-to-hexagonal transition in block copolymers and charged systems~\cite{Ohta1986,luo2025theory}.

Since the two reactions are independent, we suspect that each morphology can also be tuned independently.
To test this, we fix $r_2=3$ (so that the $CD$ phase is hexagonal) and only vary $r_1$.
\figref{fig:figure3}b shows that in this case the $AB$ phase transitions through lamellar, hexagonal, and homogeneous states while the $CD$ phase remains hexagonal.
The critical values of $r_1$, at which transitions occur, are nearly independent of $r_2$, confirming that the two charge flavors act as independent control parameters.

We finally test whether we can tune morphologies of all phases by varying $r_1$ and $r_2$ independently.
The corresponding state diagram should have $3{\times}3{=}9$ distinct two-phase states corresponding to combinations of lamellar, hexagonal, and homogeneous phases (\figref{fig:figure3}c).
Indeed, \figref{fig:figure3}d shows examples of equilibrium profiles for the four states in which both phases are patterned, confirming that all combination are attainable. %
Moreover, two-dimensional dynamical simulations of \Eqref{eqn:pde} at corresponding parameters reproduce the predicted morphologies (\figref{fig:figure3}e); stripes and droplets are clearly visible, even though the simulations have not yet reached stationary state.
Taken together, this analysis confirms that interpreting the Lyapunov-functional~$\mathcal L$ as a free energy correctly predicts the dynamical steady states of chemically active mixtures.

\subsectionCustom{Extension to many components}

The modular structure of our framework, which relies on one independent reaction per charge flavor, makes extensions to more complex systems straightforward.
In the simplest case, each additional reaction adds two new species, introduces one new neutrality constraint, and thus leads to a new patterned phase whose morphology is set by the corresponding rate ratio~$r_i$.

To demonstrate designed morphologies, we construct a six-component system ($N{=}6$, $S{=}3$, $P_{\max}{=}3$) with three reaction pairs $A\react B$, $C\react D$, and $E\react F$.
For simplicity, we use a free energy akin to \Eqref{eqn:free_energy} where all six species repel each other with equal strength $\chi$ (see \Figref{fig:snapshot_6comp} for full parameters).
Based on the analysis of the reaction rate ratios~$r_i$ above, we choose $r_1=r_2=3$ and $r_3=1$ to create two hexagonal and one lamellar phase.
Indeed, dynamical simulation of \Eqref{eqn:pde} produces the complex multiscale pattern shown in \figref{fig:snapshot_6comp}a, where droplets of two distinct sizes coexist with lamellar stripes — a seemingly intricate structure involving multiple length scales.
Numerical minimization of the associated Lyapunov function $\lyapunov$ reveals that this pattern decomposes into exactly three coexisting patterned phases (\figref{fig:snapshot_6comp}b--d): the hexagonal phase~1 enriches $A$ and $B$ ($r_1=3$), the hexagonal phase~2 enriches $C$ and $D$ forming droplets of a different size ($r_2=3$), and the lamellar phase~3 enriches $E$ and $F$ ($r_3=1$).
Each morphology is independently governed by the corresponding rate ratio~$r_i$, exactly as established in the four-component model.
This equilibrium analysis suggests that the transient state shown in \figref{fig:snapshot_6comp}a comprises three (patterned) phases, which are still coarsening. 
Indeed, each interface between two phases is equipped with a surface tension (see Appendix \ref{sec:surface} and \cite{xu2017computing,netz1997interfaces,weith2013stability}), which will drive a coarsening process akin to Ostwald ripening, so that the final stationary state will comprise the three phases shown in \figref{fig:snapshot_6comp}b--d.

\sectionCustom{Discussion and conclusion}

Our results establish a constructive design principle: an $N$-component mixture with $S$ independent reaction pairs exhibits up to $N{-}S$ coexisting phases.
The patterns' morphologies are directly controlled by the reaction rate ratios, which can be interpreted as charge asymmetry. %
The role of the generalized charges is less transparent in cases where a component is involved in more than one reaction.
In such cases, phases can exhibit more complex patterns, such as square lattices~(see \figref{fig:square_lattice}), but the generalization of Gibbs phase rule, given by \Eqref{eqn:phase_rule}, remains valid.
The phase rule only breaks down when it becomes impossible to identify the charges according to \Eqref{eqn:decomposition}, in which case traveling waves, oscillatory dynamics, and chaos become possible~\cite{Luo2023}.

Our framework complements linear stability analysis, which captures only the onset of instabilities.
Essentially, \Eqref{eqn:pde} describes non-linear spatial fluxes and linearized reactions (including a constant term in the reactions, see Appendix \ref{sec:dynamics}).
This approach allowed us to identify globally stable states via minimizing the Lyapunov functional~$\mathcal L$, and we identified coexistence of multiple patterned phases that may be missed when the homogeneous state is itself locally stable.
Most importantly, our approach identifies chemical reactions with long-range interactions, which enables us to use tools from equilibrium physics.
In the future, we could thus use ideas from crystals and nematics to investigate the interface of phases with hexagonal and lamella patterns in more detail.
Moreover, reactions might also give rise to multiple competing length scales, which could lead to quasicrystalline order~\cite{archer2015soft, archer2013quasicrystalline, duan2018stability}.
In fact, multiphase coexistence has been observed in many recent simulations and experiments~\cite{jacobs2023theory,rossetto2025exchange,greve2025coexistence}, and potentially emerges in charged polyelectrolyte systems \cite{Agrawal2025} as well as ecological communities \cite{zlrz-zvqk,alhiyasat2026spatiotemporal}.
We hope that mapping such systems to phase coexistence enables deeper understanding of spatial organization in living cells and to engineer synthetic active materials with prescribed multiscale morphologies.

\emph{Acknowledgments}---%
We thank Uwe Thiele, Filipe Thewes, and Mengmeng Wu for helpful discussions.
CL thanks Jiayu Xie, Xinxiang Chen, Shiling Liang for helpful discussions.
We gratefully acknowledge funding from the Max Planck Society and the European Research Council (ERC, EmulSim, 101044662).

\bibliographystyle{apsrev4-2}
\bibliography{main}

\newpage

\onecolumngrid
\appendix
  \renewcommand{\thefigure}{S\arabic{figure}}
  \setcounter{figure}{0}

\section{Dynamics of multicomponent mixtures with multiple long-range interactions}
\label{sec:dynamics}
We consider a $d$-dimensional incompressible, isothermal mixture composed of $\Nc$ components with charge density $q_i^{(s)}$ for $i=1,...,\Nc$, where $s=1,...,\Nr$ denotes the flavor of the charge. %
The dimensionless free energy density $\hat{f}[\{\phi_i(\br)\},\{\psi_s(\br)\},\xi(\br),\gamma_i]$ is given by
\begin{equation}
\label{eq:hatf}
    	\hat{f}=  \hat{f}_\mathrm{local}+\hat{f}_\mathrm{interface}+\hat{f}_\mathrm{long-range}+\hat{f}_\mathrm{incompressible}+\hat{f}_\mathrm{conserve}+\hat{f}_\mathrm{charge-constraint}
 \;,
\end{equation}
where 
\begin{subequations}
\begin{align}
    \label{eq:hatf_local}
    \hat{f}_\mathrm{local}&= \frac1{V}\int_V \bigg[\sum_i\frac{\phi_i(\br)}{l_i} \ln \phi_i(\br)+ \frac12\sum_{ij} \chi_{ij}\phi_i(\br)\phi_j(\br)\bigg] \mathrm{d}\vect r,
    \\
    \hat{f}_\mathrm{interface}&= \frac1{V}\int_V \bigg[\frac12\sum_i{\kappa}_{i}  |\nabla \phi_i(\br)|^2\bigg] \mathrm{d}\vect r,  
    \\
    \hat{f}_\mathrm{long-range}&= \frac1{V}\sum_s\int_V
    \bigg[-\frac12|\nabla\psi_s(\br)|^2+\psi_s(\br)\sum_iq_i^{(s)}(\phi_i(\br)-\phi_i^0)\bigg] \mathrm{d}\vect r,
    \\
    \hat{f}_\mathrm{incompressible}&= \frac1{V}\int_V \xi(\br) \bigg(\sum_{i=1}^{\Nc}\phi_i(\br)-1\bigg) \mathrm{d}\vect r,
    \\
    \hat{f}_\mathrm{conserve}&= \sum_i \gamma_i \bigg(\frac1{V}\int_V \phi_i(\br) \mathrm{d}\vect r -\bar{\phi}_i\bigg),
    \\
    \label{eq:hatf_charge_constraint}
    \hat{f}_\mathrm{charge-constraint}&=\sum_i \tilde{\zeta}_i k_{ij}\frac1{V}\int_V (\phi_j(\br)-\phi_j^0) \mathrm{d}\vect r= \sum_{s} \zeta_s \sum_i \left(q_i^{(s)} \frac1{V}\int_V  (\phi_i(\br)-\phi_i^0) \mathrm{d}\vect r\right),
\end{align}
\end{subequations}
are the local free energy, interfacial energy, long-range interaction energy, incompressibility constraint, conservation of the total volume fraction of each component, and 
charge-constraint terms, respectively. Here, $\phi_i(\br)$ is the volume fraction of component $i$ at position $\br$, $l_i$ is the dimensionless molecular size of component $i$, $\chi_{ij}$ is the Flory interaction parameter between component $i$ and $j$, ${\kappa}_{i}$ is the gradient energy coefficient of component $i$, $\psi_s(\br)$ is the dimensionless electrostatic potential field for flavor $s$, $\xi(\br)$ is the Lagrange multiplier field to enforce incompressibility, $\gamma_i$ is the Lagrange multiplier to enforce conservation of the total volume fraction of component $i$, and $\zeta_s$ is the effective Lagrange multiplier to enforce charge constraints, connecting to the original multiplier $\tilde{\zeta}_i$ via $\zeta_s=-\sum_{il} \tilde{\zeta}_i\Lambda_{il} q_l^{(s)}$ . The total volume is $V$ and $\bar{\phi}_i$ is the mean volume fraction of component $i$.
At equilibrium, the free energy is minimized with respect to $\phi_i(\br)$ and $\psi_s(\br)$, while maximized with respect to the Lagrange multipliers $\xi(\br)$, $\gamma_i$, and $\zeta_i$. 

The minimization conditions give
\begin{subequations}
\begin{align}
    0&=\frac{\delta \hat{f}}{\delta \phi_i(\br)}= \frac{1}{l_i} \ln \phi_i(\br)+ 
    \frac 1{l_i}
     + \sum_{j} \chi_{ij}\phi_j(\br)-{\kappa}_{i}  \nabla ^2 \phi_i(\br)   
    + \sum_s
    q_i^{(s)}\psi_s(\br)
    +\xi(\br)+\gamma_i+\sum_s q_i^{(s)} \zeta_s
    \nonumber \\
    &\equiv \hat{\mu}_i(\br)
    \equiv \tilde{\mu}_i(\br)+\xi(\br)+\gamma_i+\sum_s q_i^{(s)} \zeta_s
    \equiv \mu_i(\br)+\sum_s
    q_i^{(s)}\psi_s(\br)+\xi(\br)+\gamma_i+\sum_s q_i^{(s)} \zeta_s,
    \\
    0&=\frac{\delta \hat{f}}{\delta \psi_s(\br)}= \nabla^2\psi_s(\br)+\sum_iq_i^{(s)}(\phi_i(\br)-\phi_i^0),%
    \\
    0&=\frac{\delta \hat{f}}{\delta \xi(\br)}= \sum_{i=1}^{\Nc}\phi_i(\br)-1,
    \\
    0&=\frac{\partial \hat{f}}{\partial \gamma_i}= \frac1{V}\int_V \phi_i(\br) \mathrm{d}\vect r -\bar{\phi}_i,
    \\
    0&=\frac{\partial \hat{f}}{\partial \zeta_s}= \sum_i q_i^{(s)} \frac1{V}\int_V  (\phi_i(\br)-\phi_i^0) \mathrm{d}\vect r \;.
\end{align}
\end{subequations}
Note that the second equation is just Poisson's equation for each flavor of the long-range interaction.

We first show that the model B dynamics of a system with this free energy is equivalent to the dynamics with linear chemical reactions without long-range interactions.
The dynamics of $\phi_i(\br)$ is given by
\begin{equation}
    \frac{\partial \phi_i(\br)}{\partial t}=\sum_j \nabla\cdot \mathrm{J}_j=\sum_j \nabla \cdot M_{ij} \nabla \hat{\mu}_j(\br)
    =\sum_j \nabla \cdot M_{ij} \nabla \big[\tilde{\mu}_j(\br)+\xi(\br)\big]\;.
\end{equation}
Using the incompressibility condition $\sum_{i=1}^{\Nc}\phi_i(\br)=1$, we have $\sum_{i=1}^{\Nc}\partial \phi_i(\br)/\partial t=0$, thus the total flux vanishes, i.e., $\sum_{i=1}^{\Nc}\mathrm{J}_i=0$. This gives
\begin{equation}
    \nabla \xi(\br)=-\frac{\sum_{ij} M_{ij} \nabla \tilde{\mu}_j(\br)}{\sum_{ij} M_{ij}}\;.    
\end{equation}
Defining an effective mobility matrix $\Lambda_{ij}=M_{ij}-\frac{\sum_k M_{ik}\sum_l M_{lj}}{\sum_{kl} M_{kl}}$, the dynamic equations then read
\begin{eqnarray}
    \frac{\partial \phi_i(\br)}{\partial t}
    &=&\sum_j \nabla \cdot \Lambda_{ij} \nabla \tilde{\mu}_j(\br)
    =\sum_j \nabla \cdot \Lambda_{ij} \nabla \left[
        \mu_j(\br)+\sum_s
    q_j^{(s)}\psi_s(\br)
    \right]
    \;.
\end{eqnarray}
Assuming $\Lambda_{ij}$ does not depend on position $\br$ and using Poisson's equation, we have
\begin{eqnarray}
    \frac{\partial \phi_i(\br)}{\partial t}
    &=&\sum_j \Lambda_{ij} \nabla^2 \mu_j(\br)
    -\sum_j  \Lambda_{ij}\sum_k \sum_s q_j^{(s)} q_k^{(s)} (\phi_k(\br)-\phi_k^0)%
    \equiv\sum_j \Lambda_{ij} \nabla^2 \mu_j(\br) + \sum_j k_{ij}(\phi_j(\br)-\phi_j^0)%
    \;.
\end{eqnarray}
This is the equation we used in the main text, where the long-range interactions give rise to effective linear reaction terms with $k_{ij}=-\sum_l \mob_{il} \sum_s q_l^{(s)} q_j^{(s)}$. The condition for mass conservation $\sum_i k_{ij}=0$ is satisfied since $\sum_i \Lambda_{ij}=0$.
For simplicity, we set $\phi_i^0=0$ in the derivation below.

\section{Generalized Gibbs-ensemble method for finding coexisting phases}
\label{sec:Gibbs}

To study the coexistence of different phases, rather than minimizing the total free energy in a single big box, we consider a Gibbs ensemble of $M$ phases, where the fraction of volume of phase $\beta$ is $J_\beta$ with $\sum_{\beta=1}^M J_\beta = 1$. The total free energy density of the system is then given by
\begin{eqnarray}
    \bar{f} = \sum_{\beta=1}^M J_\beta \hat{f}^{(\beta)},
\end{eqnarray}
where
$\hat{f}^{(\beta)}$ 
is the free energy density of phase $\beta$ with volume fraction profiles $\phi_i^{(\beta)}(\br)$, electrostatic potential $\psi_s^{(\beta)}(\br)$, volume $V_\beta$ and mean volume fractions $\phibar_i^{(\beta)}$. To enforce the conservation of the total volume fraction of each component, an extra constraint is $\sum_{\beta} J_\beta \phibar_i^{(\beta)}=\bar{\phi}_i$ for each component $i$. This can be achieved by introducing single chain potential fields $\omega_i^{(\beta)}(\br)$ conjugated to $\phi_i^{(\beta)}(\br)$ and the single chain partition function $Q_i$ for each component $i$. Similarly, we can assure the constraint $\sum_{\beta=1}^M J_\beta = 1$  by introducing an extra variable $\nu_\beta$ and the corresponding partition function $P$.  Explicitly, we write a new total free energy density in the form
\begin{eqnarray}
    {f} = \sum_{\beta=1}^M J_\beta {f}^{(\beta)}-\sum_i \frac{\bar{\phi}_i}{l_i}\ln Q_i - \ln P-\sum_\beta J_\beta\nu_\beta-\sum_\beta J_\beta\ln J_\beta,
\end{eqnarray}
where the single chain partition function $Q_i$ is defined as
\begin{eqnarray}
    \label{eq:Qi}
    Q_i=\sum_\beta J_\beta\frac1{\Vbeta}\int_{\Vbeta} e^{-\omega_i^{(\beta)}(\br) l_i}\mathrm{d}\vect r,
\end{eqnarray}
and 
\begin{eqnarray}
    \label{eq:Pi}
    P=\sum_\beta e^{-\nu_\beta}.
\end{eqnarray}
Moreover, ${f}^{(\beta)}$ 
is the free energy density of phase $\beta$ in a box with length $L_{\beta}^j$ in the direction $\br_j$ ($j=1,\hdots,d$), with volume fraction fields $\phi_i^{(\beta)}(\br)$, single chain potential fields $\omega_i^{(\beta)}(\br)$, long-range potentials $\psi_s^{(\beta)}(\br)$, and volume $V_\beta=\prod_j L_{\beta}^j$.
Each phase's free energy density has five parts, 
\begin{eqnarray}
    {f}^{(\beta)} = {f}_\mathrm{local}^{(\beta)} + {f}_\mathrm{interface}^{(\beta)} + {f}_\mathrm{long-range}^{(\beta)}+ {f}_\mathrm{incompressible}^{(\beta)} + {f}_\mathrm{charge-constraint}^{(\beta)}\;.
\end{eqnarray}
Explicitly, these terms read
\begin{subequations}
    \begin{align}
        {f}_\mathrm{local}^{(\beta)}&= \frac1{V_\beta}\int_{V_\beta} \bigg[-\sum_i \phi_i^{(\beta)}(\br)\omega_i^{(\beta)}(\br)+ \frac12\sum_{ij} \chi_{ij}\phi_i^{(\beta)}(\br)\phi_j^{(\beta)}(\br)\bigg] \mathrm{d}\vect r,
        \\
        {f}_\mathrm{interface}^{(\beta)}&= \frac1{V_\beta}\int_{V_\beta} \bigg[\frac12\sum_i{\kappa}_{i}  |\nabla \phi_i^{(\beta)}(\br)|^2\bigg] \mathrm{d}\vect r,  
        \\
        {f}_\mathrm{long-range}^{(\beta)}&= \frac1{V_\beta}\sum_s\int_{V_\beta}
        \bigg[-\frac12|\nabla\psi_s^{(\beta)}(\br)|^2+\psi_s^{(\beta)}(\br)\sum_iq_i^{(s)}\phi_i^{(\beta)}(\br)\bigg] \mathrm{d}\vect r,
        \\
        {f}_\mathrm{incompressible}^{(\beta)}&= \frac1{V_\beta}\int_{V_\beta} \xi^{(\beta)}(\br) \bigg(\sum_{i=1}^{\Nc}\phi_i^{(\beta)}(\br)-1\bigg) \mathrm{d}\vect r,
        \\
        {f}_\mathrm{charge-constraint}^{(\beta)}&= \sum_s\zeta_s^{(\beta)}\left(\frac1 {\Vbeta} \int_{\Vbeta}\sum_i z^s_i\phi_i^{(\beta)}(\br) \mathrm{d}\vect r\right),
    \end{align}
\end{subequations}
    where $\xi^{(\beta)}(\br)$ and $\zeta_i^{(\beta)}$ are Lagrange multipliers to enforce incompressibility and constraints such as charge neutrality in phase $\beta$, respectively.
 Note that for a thermodynamically large system, interfaces between phases do not influence the bulk properties of the phases at equilibrium, so we do not consider them.
 Below, we will see that the minimum of ${f}$ automatically satisfies the mass conservation thanks to the presence of $Q_i$, thus we do not need to enforce mass conservation explicitly here.
 The total free energy density depends on the single chain potentials, 
 the volume fraction fields, the electrostatic potential field, the volumes, fractions of volumes for each phase, and the Lagrange multipliers, i.e., $\bar{f}= \bar{f}(\{\omega_i^{(\beta)}(\br)\}, \{
 \phi_i^{(\beta)}(\br)\}, \{\psi^{(\beta)}(\br)\}, \{V_\beta\}, \{J_\beta\},\{\xi^{(\beta)}(\br)\}, \{\zeta_s^{(\beta)}\},\{\nu_\beta\}
 )$. 
The equilibrium coexisting states can be obtained by minimizing $f$ 
over all these fields and variables. 
The extremum of the free energy density $f$ with respect to $\omega_i^{(\beta)}(\br)$ gives
\begin{equation}
	\label{eq:phi_i_r_beta}
	\phi_i^{(\beta)}(\br)={\phibar_i}\frac{e^{-\omega_i^{(\beta)}(\br) l_i}}{Q_i}
    \;.
\end{equation}
Note that this automatically ensures material conservation,
\begin{equation}
	 \sum_{\beta=1}^{M} J_{\beta}\frac1\Vbeta\int_\Vbeta\phi_i^{(\beta)}(\br)\mathrm{d}\vect r ={\phibar_i}
  \;.
\end{equation}
Inserting \Eqref{eq:phi_i_r_beta} into the free energy density $f$, we can see that the term involving $\omega_i^{(\beta)}(\br)$ gives rise to an entropy contribution
 $\sum_i\frac{\phi_i^{(\beta)}(\br)}{l_i} \ln \phi_i^{(\beta)}(\br)$, which is the same as the entropy term in \Eqref{eq:hatf_local}, thus proving $f=\bar{f}$ when both are minimized.
The extremum of $f$ with respect to  
$\phi_i^{(\beta)}(\br)$  provides
 \begin{equation}
 \label{eq:omega_i_r_beta}
	\omega_i^{(\beta)}(\br)=\sum_j \chi_{ij}\phi_j^{(\beta)}(\br)-{\kappa}_i\nabla^2\phi_i^{(\beta)}(\br)+\xi^{(\beta)}(\br)+\sum_sq_i^{(s)}\psi_s^{(\beta)}(\br)+\sum_sq_i^{(s)}\zeta_s^{(\beta)}
 \;.
 \end{equation}
The extremum of $f$ with respect to  $\psi_s^{(\beta)}(\br)$ gives Poisson's equation
 \begin{equation}
 \label{eq:poisson}
	\nabla^2\psi_s^{(\beta)}(\br)=-\sum_iq_i^{(s)}\phi_i^{(\beta)}(\br)
 \;.
 \end{equation} 
The extremum of $f$ with respect to $J_\beta$ provides
\begin{align}
    \label{eq:nubeta}
    \nu_\beta=f^{(\beta)}-\ln J_\beta-1-\sum_i \frac1{l_i} \frac1{\Vbeta}\int_\Vbeta \phi_i(\beta,\br)\mathrm{d}\vect r
    \;.
\end{align}
The extremum of $f$ with respect to $\nu_\beta$ provides
\begin{align}
    \label{eq:Jbeta}
   J_\beta=\frac1Pe^{-\nu_\beta}
    \;.
\end{align}
The extremum of $f$ with respect to $\zeta_s^{(\beta)}$ gives charge constraints in each compartment,
\begin{equation}
\label{eq:neutral}
	\frac1{\Vbeta} \int_{\Vbeta}\sum_i q_i^{(s)}\phi_i(\beta,\br) \mathrm{d}\vect r =0
 \;.
\end{equation}
The extremum of $f$ with respect to $\xi(\beta,\br)$ simply gives incompressibility in each compartment,
 \begin{equation}
 \label{eq:incomp}
	\sum_i\phi_i^{(\beta)}(\br)=1\;.
 \end{equation}
    Finally,
the extremum of $f$ with respect to $\Lbeta^j$ gives
\begin{align}
	\frac{\mathrm{d} f^*}{\mathrm{d} \Lbeta^j}
 &=\left.\frac{\partial f}{\partial  \Lbeta^j}\right|_*
 +
 \int \sum_i \left.\frac{\delta f}{\delta \phi_i}\right\vert_*\frac{\partial \phi_i^*}{\partial\Lbeta^j}\mathrm{d}\br
	+\int \sum_i \left.\frac{\delta f}{\delta w_i}\right\vert_*\frac{\partial w_i^*}{\partial \Lbeta^j}\mathrm{d}\br
	+\int  \left.\frac{\delta f}{\delta\psi}\right\vert_*\frac{\partial \psi^*}{\partial \Lbeta^j}\mathrm{d}\br
	+\int  \left.\frac{\delta f}{\delta\xi}\right\vert_*\frac{\partial \xi^*}{\partial \Lbeta^j}\mathrm{d}\br+...
	\notag \\
	&=\left.\frac{\partial f}{\partial  \Lbeta^j}\right|_*
	\notag \\
	&=\frac{\partial}{\partial \Lbeta^j} \left[{J_\beta}\frac{1}{\Vbeta} \int_\Vbeta\biggl( \frac12\sum_i{\kappa}_{i}  |\nabla \phi_i(\beta,\br)|^2
	-\sum_s\frac12|\nabla\psi_s(\beta,\br)|^2\biggr) \mathrm{d}\br\right]_*
 =0\;,
 \end{align}
 where the star denotes that quantities are evaluated for profiles that have been obtained by optimizing over all fields and parameters except $\Lbeta^j$.
 In particular, $f^*(\Lbeta^j)$ denotes the associated free energy density, which then only depends on $\Lbeta^j$.
 Using $f_\mathrm{int}^j=\Vbeta^{-1} \int_\Vbeta \frac12\sum_i{\kappa}_{i}  (\partial_{j} \phi_i(\beta,\br))^2 \mathrm{d}\br$ and $f_\mathrm{\psi}^j=-\Vbeta^{-1} \int_\Vbeta \sum_s\frac12(\partial_j\psi_s(\beta,\br))^2  \mathrm{d}\br$,
  we obtain 
 \begin{align}
 \label{eq:Lbeta}
 \frac{\mathrm{d} f}{\mathrm{d} \Lbeta^j}&=-J_\beta\frac{2}{\Lbeta^j}(f_\mathrm{int}^j+f_\mathrm{\psi}^j)=0\;,
  \end{align}
  which gives 
  \begin{eqnarray}
      f_\mathrm{int}^j=-f_\mathrm{\psi}^j\;.
  \end{eqnarray}
  Using Poisson's equation and integration by parts, we find
  \begin{eqnarray}
    \label{eqn:equivalence_electric_interface}
      f_{\mathrm{el}}=\frac{1}{\Vbeta} \int_\Vbeta\sum_s \bigg[-\frac12(\nabla\psi_s(\beta,\br))^2  +\psi_s(\beta,\br) \sum_i q_i^{(s)}\phi_i(\beta,\br)\bigg] \mathrm{d}\br=-\sum_j f_\mathrm{\psi}^j,
  \end{eqnarray}
  and thus $f_{\mathrm{el}}=f_\mathrm{int}$ for all compartments, where $f_\mathrm{int}=\sum_j f_\mathrm{int}^j$ is the free energy density associated with interfaces.
  This indicates that \textit{the total long-range energy over all flavors is balanced by the interfacial energy in equilibrium.}
 Specifically, in the 1D case, we have
 \begin{eqnarray}
 \label{eq:Lbeta_1D}
	\frac{\mathrm{d} f}{\mathrm{d} \Lbeta}&=&-{J_\beta}\frac{2}{\Lbeta} \frac1\Lbeta\int_0^\Lbeta\Bigg[ \frac12\sum_i{\kappa}_{i}  |\partial_x \phi_i(\beta,x)|^2
	-\sum_s\frac12|\partial_x\psi_s(\beta,x)|^2\bigg] \mathrm{d}x 
	\nonumber\\
	&=&-{J_\beta}\frac{2}{\Lbeta} \frac1\Lbeta\int_0^\Lbeta\Bigg[ -\frac12\sum_i{\kappa}_{i}  \phi_i(\beta,x)\partial_x^2 \phi_i(\beta,x)
	+\sum_s\frac12\psi_s(\beta,x)\partial_x^2\psi_s(\beta,x)\bigg] \mathrm{d}x =0
 \;.
\end{eqnarray}
In the 2D case, we have
\begin{subequations}
    \begin{eqnarray}
        \label{eq:Lbeta_2Dx}
        \frac{\mathrm{d} f}{\mathrm{d} \Lbeta^x}&=&-{J_\beta}\frac{2}{\Lbeta^x} \frac1{\Lbeta^x\Lbeta^y}\int_0^{\Lbeta^x}\int_0^{\Lbeta^y}\Bigg[ \frac12\sum_i{\kappa}_{i}  |\partial_x \phi_i(\beta,x,y)|^2
        -\sum_s\frac12|\partial_x\psi_s(\beta,x,y)|^2\bigg] \mathrm{d}x \mathrm{d}y
        \nonumber\\
        &=&-{J_\beta}\frac{2}{\Lbeta^x} \frac1{\Lbeta^x\Lbeta^y}\int_0^{\Lbeta^x}\int_0^{\Lbeta^y}\Bigg[ -\frac12\sum_i{\kappa}_{i}  \phi_i(\beta,x,y)\partial_x^2 \phi_i(\beta,x,y)
        +\sum_s\frac12\psi_s(\beta,x,y)\partial_x^2\psi_s(\beta,x,y)\bigg] \mathrm{d}x\mathrm{d}y 
        \nonumber\\
        &=&0
        \;.
    \end{eqnarray}
    \begin{eqnarray}
    \label{eq:Lbeta_2Dy}
    \frac{\mathrm{d} f}{\mathrm{d} \Lbeta^y}&=&-{J_\beta}\frac{2}{\Lbeta^y} \frac1{\Lbeta^x\Lbeta^y}\int_0^{\Lbeta^x}\int_0^{\Lbeta^y}\Bigg[ \frac12\sum_i{\kappa}_{i}  |\partial_y \phi_i(\beta,x,y)|^2
    -\sum_s\frac12|\partial_y\psi_s(\beta,x,y)|^2\bigg] \mathrm{d}x \mathrm{d}y
    \nonumber\\
    &=&-{J_\beta}\frac{2}{\Lbeta^y} \frac1{\Lbeta^x\Lbeta^y}\int_0^{\Lbeta^x}\int_0^{\Lbeta^y}\Bigg[ -\frac12\sum_i{\kappa}_{i}  \phi_i(\beta,x,y)\partial_y^2 \phi_i(\beta,x,y)
    +\sum_s\frac12\psi_s(\beta,x,y)\partial_y^2\psi_s(\beta,x,y)\bigg] \mathrm{d}x\mathrm{d}y 
    \nonumber\\
    &=&0
    \;.
\end{eqnarray}
\end{subequations}
In summary, we obtain the self-consistent equations \eqref{eq:Qi}, \eqref{eq:Pi}, \eqref{eq:phi_i_r_beta},
\eqref{eq:omega_i_r_beta}--\eqref{eq:incomp},
and \eqref{eq:Lbeta} to determine equilibrium states.
We numerically solve these equations using a scheme similar to the approach introduced in \cite{luo2025theory}. We use this method to obtain the coexisting phases shown in all figures in the main text.

\section{Stability analysis for multicomponent mixtures with chemical reactions}
To analyze the stability of stationary states of Eq.~(1) in the main text, we consider the Hessian matrix of the free energy density $\hat{f}$ defined in \Eqref{eq:hatf}. %
Assuming the homogeneous state $\phi_i(\br)=\bar{\phi}_i$, $\psi_s(\br)=0$, we consider small perturbations $\delta \phi_i(\br)$, $\delta \psi_s(\br)$ around this state. Expanding the free energy density up to second order in these perturbations gives
\begin{align}
    \delta^2 \hat{f}
    &= \frac1{2V}\int_V \bigg[
        \sum_i \frac{1}{l_i \bar{\phi}_i} \big(\delta \phi_i(\br)\big)^2
        + \sum_{ij} \chi_{ij}\, \delta \phi_i(\br)\,\delta \phi_j(\br)
        + \sum_i \kappa_i \big|\nabla \delta \phi_i(\br)\big|^2
    \bigg] \,\mathrm{d}\vect r
    \nonumber\\
    &\quad
    + \frac1{V}\sum_s \int_V \bigg[
        -\frac12 \big|\nabla \delta \psi_s(\br)\big|^2
        + \delta \psi_s(\br) \sum_i q_i^{(s)} \,\delta \phi_i(\br)
    \bigg] \,\mathrm{d}\vect r
    \;.
\end{align}
Written in Fourier space, this becomes
\begin{align}
    \delta^2 \hat{f}
    &= \frac1{2V}\sum_{\vect q} \bigg[
        \sum_{ij}
        \bigg(
            \frac{1}{l_i \bar{\phi}_i} \delta_{ij}
            + \chi_{ij}
            + \kappa_i q^2 \delta_{ij}
        \bigg)
        \delta \phi_i(-\vect q)\, \delta \phi_j(\vect q)
    \bigg]
    \nonumber\\
    &\quad
    + \frac1{V}\sum_{\vect q} \sum_s \bigg  [
        \frac12 |\vect q|^2 \big|\delta \psi_s(\vect q)\big|^2
        + \delta \psi_s(-\vect q) \sum_i q_i^{(s)} \,\delta \phi_i(\vect q)
    \bigg]
    \;.
\end{align}
Using the Poisson equation in Fourier space, $|\vect q|^2 \delta \psi_s(\vect q) = \sum_i q_i^{(s)} \,\delta \phi_i(\vect q)$, we can eliminate $\delta \psi_s(\vect q)$ to obtain
\begin{eqnarray}
    \delta^2 \hat{f}
    &=& \frac1{2V}\sum_{\vect q} \bigg[
        \sum_{ij}
        \bigg(
            \frac{1}{l_i \bar{\phi}_i} \delta_{ij}
            + \chi_{ij}
            + \kappa_i |\vect q|^2 \delta_{ij}
            + \sum_s \frac{q_i^{(s)} q_j^{(s)}}{|\vect q|^2}
        \bigg)
        \delta \phi_i(-\vect q)\, \delta \phi_j(\vect q)
    \bigg]
    \;.
\end{eqnarray}
Thus, the Hessian matrix of the homogeneous state in Fourier space is given by
\begin{eqnarray}
    \mathcal{H}_{ij}(\vect q)
    &=& \frac{1}{l_i \bar{\phi}_i} \delta_{ij}
        + \chi_{ij}
        + \kappa_i q^2 \delta_{ij}
        + \sum_s \frac{q_i^{(s)} q_j^{(s)}}{q^2}
    \;.
\end{eqnarray}
We further use incompressibility to eliminate one component, e.g., component $\Nc$, leading to an effective Hessian matrix for the remaining $\Nc-1$ components,
\begin{eqnarray}
    \mathcal{H}_{ij}^\mathrm{eff}(\vect q)
    &=& \mathcal{H}_{ij}(\vect q)
        - \mathcal{H}_{i\Nc}(\vect q)
        - \mathcal{H}_{\Nc j}(\vect q)
        + \mathcal{H}_{\Nc \Nc}(\vect q)
    \;,
\end{eqnarray}
for $i,j=1,\hdots,\Nc-1$.
The homogeneous state is linearly unstable if there exists a wave vector $\vect q$ such that $\mathcal{H}^\mathrm{eff}(\vect q)$ has a negative eigenvalue. The spinodal curves shown in Fig. 2 in the main text are obtained using this stability analysis.

\section{Surface tensions between coexisting phases}
\label{sec:surface}
To calculate the surface tension between two coexisting phases, we consider a grand canonical ensemble.
At fixed chemical potentials $\mu_i=\mu_i^*$, we can obtain (i) the equilibrium profiles of volume fractions $\phi_i(x)$ and potentials $\psi_s(x)$ with an interface between the two phases using the dynamic equations, and (ii) the equilibrium profiles of each phase without interface. The grand potential density is given by
\begin{eqnarray}
    \omega(x)=f(x)-\sum_i \mu_i^* \phi_i(x)
    \;.
\end{eqnarray}
The grand potential densities of the two coexisting phases without interface must be equal, denoted as $\omega_\mathrm{phase1}$ and $\omega_\mathrm{phase2}$.
The surface tension is then given by
\begin{eqnarray}
    \gamma=\int_{-\infty}^{\infty} \big[\omega(x)-\omega_\mathrm{phase1}\big] \mathrm{d}x
    \;.
\end{eqnarray}
In fact, for the coexistence of two homogenenous phase in 1D, the surface tension $\gamma$ can be calculated using the simple formula only using the equilibrium profiles with interface:
\begin{eqnarray}
    \gamma=\int_{-\infty}^{\infty} \bigg[\sum_i \kappa_i \left(\frac{\mathrm{d}\phi_i(x)}{\mathrm{d}x}\right)^2 -\sum_s \left(\frac{\mathrm{d}\psi_s(x)}{\mathrm{d}x}\right)^2 \bigg] \mathrm{d}x
    \;.
\end{eqnarray}
To derive this equation, we consider the free energy density without interfacial energy and electrostatic energy, $f_\mathrm{L}(\{\phi_i\})$, and rewrite the grand potential as
\begin{eqnarray}
    \Omega=\int_{-\infty}^{\infty} dx
    [f_\mathrm{L}(\{\phi_i(x)\})+\sum_i \frac12 \kappa_i \big(\frac{\mathrm{d}\phi_i(x)}{\mathrm{d}x}\big)^2 -\sum_s \frac12 \big(\frac{\mathrm{d}\psi_s(x)}{\mathrm{d}x}\big)^2 + \sum_s \sum_i q_i^{(s)} \psi_s(x) \phi_i(x)
    - \sum_i \mu_i^* \phi_i(x)]
    \;.
\end{eqnarray}
Using the Euler-Lagrange equation for $\phi_i(x)$ and $\psi_s(x)$, we have
\begin{subequations}
    \begin{align}
        \frac{\delta \Omega}{\delta \phi_i(x)}=0
        &\Rightarrow
        \frac{\partial f_\mathrm{L}}{\partial \phi_i} -\kappa_i \frac{\mathrm{d}^2 \phi_i(x)}{\mathrm{d}x^2} + \sum_s q_i^{(s)} \psi_s(x)=\mu_i^*
        \;,\\
        \frac{\delta \Omega}{\delta \psi_s(x)}=0
        &\Rightarrow
        \frac{\mathrm{d}^2 \psi_s(x)}{\mathrm{d}x^2} =- \sum_i q_i^{(s)} \phi_i(x)
        \;.
    \end{align}
\end{subequations}
Multiplying the first equation by $\frac{\mathrm{d}\phi_i(x)}{\mathrm{d}x}$ and the second equation by $\frac{\mathrm{d}\psi_s(x)}{\mathrm{d}x}$, and summing over $i$ and $s$, we obtain
\begin{subequations}
    \begin{align}
        \sum_i \frac{\partial f_\mathrm{L}}{\partial \phi_i} \frac{\mathrm{d}\phi_i(x)}{\mathrm{d}x}
        &= \sum_i \kappa_i \frac{\mathrm{d}^2 \phi_i(x)}{\mathrm{d}x^2} \frac{\mathrm{d}\phi_i(x)}{\mathrm{d}x}
        - \sum_s \sum_i q_i^{(s)} \psi_s(x) \frac{\mathrm{d}\phi_i(x)}{\mathrm{d}x}
        + \sum_i \mu_i^* \frac{\mathrm{d}\phi_i(x)}{\mathrm{d}x}
        \;,\\
        \sum_s \frac{\mathrm{d}^2 \psi_s(x)}{\mathrm{d}x^2} \frac{\mathrm{d}\psi_s(x)}{\mathrm{d}x}
        &= -\sum_s \sum_i q_i^{(s)} \phi_i(x) \frac{\mathrm{d}\psi_s(x)}{\mathrm{d}x}
        \;.
    \end{align}
\end{subequations}
Adding these two equations gives
\begin{eqnarray}
    \frac{\mathrm{d}}{\mathrm{d}x} \bigg[f_\mathrm{L}(\{\phi_i(x)\}) - \sum_i \mu_i^* \phi_i(x)
    - \sum_i \frac12 \kappa_i \big(\frac{\mathrm{d}\phi_i(x)}{\mathrm{d}x}\big)^2
    + \sum_s \frac12 \big(\frac{\mathrm{d}\psi_s(x)}{\mathrm{d}x}\big)^2
    + \sum_s \sum_i q_i^{(s)} \psi_s(x) \phi_i(x)
     \bigg]=0
    \;.
\end{eqnarray}
This indicates that the quantity in the square bracket is a constant, which can be determined by considering the limit $x\to \pm \infty$.
We thus have
\begin{multline}
    f_\mathrm{L}(\{\phi_i(x)\}) - \sum_i \mu_i^* \phi_i(x)
    - \sum_i \frac12 \kappa_i \big(\frac{\mathrm{d}\phi_i(x)}{\mathrm{d}x}\big)^2
    + \sum_s \frac12 \big(\frac{\mathrm{d}\psi_s(x)}{\mathrm{d}x}\big)^2
    + \sum_s \sum_i q_i^{(s)} \psi_s(x) \phi_i(x)
    \nonumber \\
    =  \left(f_\mathrm{L}(\{\phi_i(x)\}) - \sum_i \mu_i^* \phi_i(x)\right)|_{\phi_i=\phi_i^{H_B}}=
    \omega_{H_B}=
    \omega_{H_C}
    \;.
\end{multline}
Rearranging this equation gives
\begin{eqnarray}
    \omega(x)
    -\omega_{H_B}
    = \sum_i \frac12 \kappa_i \big(\frac{\mathrm{d}\phi_i(x)}{\mathrm{d}x}\big)^2
    - \sum_s \frac12 \big(\frac{\mathrm{d}\psi_s(x)}{\mathrm{d}x}\big)^2
    \;.
\end{eqnarray}
Thus, the surface tension can be written as
\begin{eqnarray}
    \gamma
    &=&\int_{-\infty}^{\infty} \big[\omega(x)-\omega_{H_B}\big] \mathrm{d}x
    \nonumber\\
    &=&\int_{-\infty}^{\infty} \bigg[\sum_i \frac12 \kappa_i \big(\frac{\mathrm{d}\phi_i(x)}{\mathrm{d}x}\big)^2
    - \sum_s \frac12 \big(\frac{\mathrm{d}\psi_s(x)}{\mathrm{d}x}\big)^2 \bigg] \mathrm{d}x
    \;.
\end{eqnarray}
It is clear that compared to passive systems, long-range contributions reduces the surface tension.
The surface tension between two patterned phases can be numerically calculated using the general formula based on grand potential densities as described above, but results will be shown elsewhere. The method is similar to the one used in \cite{netz1997interfaces}.

\section{Derivation of linear reactions from fundamental reactions}
Consider fundamental reactions of two species $A_i$ and $A_j$ with reaction rates $k_{ij}$ (from j to i) and $k_{ji}$ (from i to j):
\begin{eqnarray}
    A_i \rightleftharpoons A_j 
    \;.
\end{eqnarray}
The reaction rates of species $A_i$ and $A_j$ are given by
\begin{subequations}
    \begin{align}
        R_i(\br)&=k_{ij}\phi_j(\br)-k_{ji}\phi_i(\br)
        \;,\\
        R_j(\br)&=k_{ji}\phi_i(\br)-k_{ij}\phi_j(\br)
        \;.
    \end{align}
\end{subequations}
For a system with $\Nc$ components, we can have multiple fundamental reactions between different pairs of components. The reaction rate of component $A_i$ is given by
\begin{eqnarray}
    R_i(\br)=\sum_{j\neq i} \left[k_{ij}\phi_j(\br)-k_{ji}\phi_i(\br)\right]
    \;.
\end{eqnarray}
We can rewrite the reaction rates in the form of linear reactions as
\begin{eqnarray}
    R_i(\br)=\sum_{j}k_{ij}\phi_j(\br)
    \;,
\end{eqnarray}
where
\begin{eqnarray}
    k_{ij}=\begin{cases}
        k_{ij}, & i\neq j \\
        -\sum_{l\neq i} k_{li}, & i=j
    \end{cases}
    \;.
\end{eqnarray}
Thus, we can derive the linear reaction rates from fundamental reactions between different pairs of components.

\section{Details of mapping from reaction rates to charges}
\label{sec:mapping}
For concreteness, we assume a diagonal bare mobility matrix, $M_{ij}=\delta_{ij} m_i D_0$, where $D_0$ is a reference diffusion coefficient and $m_i$ is the dimensionless mobility of component~$i$. The reduced mobility matrix $\Lambda_{ij}$ that accounts for the no-void constraint reads~\cite{Zwicker2025} 
\begin{equation}
    \Lambda_{ij} = M_{ij} - \frac{\bigl(\sum_k M_{ik}\bigr)\bigl(\sum_l M_{lj}\bigr)}{\sum_{k,l} M_{kl}} \;.
\end{equation}
For diagonal $M_{ij}$ this simplifies to
\begin{equation}
\Lambda_{ij} = D_0 \left( m_i \delta_{ij} - \frac{m_i m_j}{\sum_k m_k} \right)
    \;.
\end{equation}
Based on these choices, we next discuss how to choose the generalized charges for $N=2$ and $N=4$ components.

\subsection{Two components}
Specializing the mobility matrix to two components $A$ and $B$, we obtain
\begin{align}
    \Lambda&=D_0\begin{pmatrix}
        m_A & 0 \\
        0 & m_B 
    \end{pmatrix}
    -
     \frac{D_0}{m_A + m_B}
    \begin{pmatrix}
       m_A^2 & m_A m_B \\
        m_A m_B & m_B^2
    \end{pmatrix}
    \nonumber \\
    &= 
    \frac{D_0 m_A m_B}{m_A + m_B}
    \begin{pmatrix}
       1 & -1 \\
        -1 & 1
    \end{pmatrix}
     \;.
\end{align}
We assume that the only flavor of effective charges is given by $q_A$ and $q_B=-q_Ar_1$. The charge matrix $Q_{ij} \equiv \sum_s q_i^{(s)} q_j^{(s)}$ then becomes
\begin{eqnarray}
    Q=\sum_sq_i^{(s)}q_j^{(s)}=q_A^2\begin{pmatrix}
        1&-r_1
        \\
        -r_1&r_1^2
    \end{pmatrix}\;.
\end{eqnarray}
The reaction matrix $K_{ij} = - \sum_k \Lambda_{ik} Q_{kj}$ becomes
\begin{align}
    K &=-\frac{D_0 m_A m_B}{m_A + m_B}q_A^2
    \begin{pmatrix}
        1+r_1& -r_1(1+r_1) \\
        -(1+r_1) & r_1(1+r_1)
    \end{pmatrix}
    \nonumber\\
    &=\frac{D_0 m_A m_B}{m_A + m_B}q_A^2( 1+r_1)
    \begin{pmatrix}
        -1& r_1 \\
        1 & -r_1.
    \end{pmatrix}
\end{align}
From the off-diagonal entries we identify
\begin{equation}
    k_{AB} = \frac{D_0 m_A m_B}{m_A + m_B} q_A^2 (1+r_1) r_1
    \qquad \text{and} \qquad
    k_{BA} = \frac{D_0 m_A m_B}{m_A + m_B} q_A^2 (1+r_1) 
    \;,
\end{equation}
so that $k_{AB}/k_{BA} = r_1$.
Inverting these relations yields the charges for given reaction rates:
\begin{align}
    q_A &= \sqrt{\frac{m_A + m_B}{D_0 m_A m_B (1+r_1) r_1}k_{AB}} 
    &
    q_B &= r_1 q_A
    \;.
\end{align}
For simplicity we set $m_A=1$ and $m_B=1/r_1$, which gives $k_{AB}=D_0r_1q_A^2$, $k_{BA}=D_0q_A^2$, or equivalently $q_A=\sqrt{\frac{k_{AB}}{D_0 r_1}}=\sqrt{\frac{k_{BA}}{D_0}}$.

\subsection{Four components}
In the case of four components, we find
\begin{align}
    \Lambda&=D_0\begin{pmatrix}
        m_A & 0 & 0 & 0\\
        0 & m_B &0 &0\\
        0& 0& m_C & 0\\
        0&0&0&m_D
    \end{pmatrix}
    -
     \frac{D_0}{m_A + m_B+m_C+m_D} \begin{pmatrix}
        m_A^2 & m_A m_B & m_A m_C & m_A m_D \\
         m_A m_B & m_B^2 &m_B m_C &m_B m_D\\
         m_A m_C &m_B m_C & m_C^2 & m_C m_D\\
         m_A m_D & m_B m_D &m_C m_D & m_D^2
     \end{pmatrix}
    \nonumber \\
    &= 
    -\frac{D_0}{m_A + m_B+m_C+m_D}
    \nonumber\\
    &\times
    \begin{pmatrix}
        -m_A (m_B + m_C + m_D) & m_A m_B & m_A m_C & m_A m_D \\
          m_A m_B & -m_B(m_A +m_C +m_D )&m_B m_C &m_B m_D\\
          m_A m_C &m_B m_C &- m_C(m_A +m_B + m_D) & m_C m_D\\
          m_A m_D & m_B m_D &m_C m_D & -m_D(m_A+ m_B + m_C )
      \end{pmatrix}
     \;.
\end{align}
To account for the two reactions $A \rightleftharpoons B$ and $C \rightleftharpoons D$, we introduce two independent flavors of charges:
$q_A^{(1)}=q_A$, $q_B^{(1)}=-q_A^{(1)}r_1$, $q_C^{(2)}=q_C$, and $q_D^{(2)}=-q_C^{(2)}r_2$.
The charge matrix $Q_{ij} = \sum_s q_i^{(s)} q_j^{(s)}$ becomes block-diagonal,
\begin{eqnarray}
    Q=\begin{pmatrix}
        q_A^2 & -q_A^2 r_1 & 0 & 0 \\
        -q_A^2 r_1 & q_A^2 r_1^2 & 0 & 0 \\
        0 & 0 & q_C^2 & -q_C^2r_2 \\
        0 & 0 & -q_C^2r_2 & q_C^2r_2^2
    \end{pmatrix}
    \;.
\end{eqnarray}
The reaction matrix $K_{ij} = - \sum_k \Lambda_{ik} Q_{kj}$ then reads
\begin{equation}
    K =\frac{D_0}{S}\begin{pmatrix} K_{AA} & K_{AB} & K_{AC} & K_{AD} \\[4pt] K_{BA} & K_{BB} & K_{BC} & K_{BD} \\[4pt] K_{CA} & K_{CB} & K_{CC} & K_{CD} \\[4pt] K_{DA} & K_{DB} & K_{DC} & K_{DD} \end{pmatrix},
\end{equation}
with $S = m_A + m_B + m_C + m_D$.
The explicit entries are
\begin{align}
    K_{AA} &= -m_A q_A^2\bigl[(1+r_1)m_B + m_C + m_D\bigr] \nonumber\\
    K_{AB} &= r_1 m_A q_A^2\bigl[(1+r_1)m_B + m_C + m_D\bigr] \nonumber\\
    K_{AC} &= m_A q_C^2(m_C - r_2 m_D) \nonumber\\
    K_{AD} &= -r_2 m_A q_C^2(m_C - r_2 m_D) \nonumber\\    
    K_{BA} &= m_B q_A^2\bigl[(1+r_1)m_A + r_1(m_C + m_D)\bigr] \nonumber\\
    K_{BB} &= -r_1 m_B q_A^2\bigl[(1+r_1)m_A + r_1(m_C + m_D)\bigr] \nonumber\\
    K_{BC} &= m_B q_C^2(m_C - r_2 m_D) \nonumber\\
    K_{BD} &= -r_2 m_B q_C^2(m_C - r_2 m_D) \nonumber\\
    K_{CA} &= m_C q_A^2(m_A - r_1 m_B) \nonumber\\
    K_{CB} &= -r_1 m_C q_A^2(m_A - r_1 m_B) \nonumber\\
    K_{CC} &= -m_C q_C^2\bigl[m_A + m_B + (1+r_2)m_D\bigr] \nonumber\\
    K_{CD} &= r_2 m_C q_C^2\bigl[m_A + m_B + (1+r_2)m_D\bigr] \nonumber\\
    K_{DA} &= m_D q_A^2(m_A - r_1 m_B) \nonumber\\
    K_{DB} &= -r_1 m_D q_A^2(m_A - r_1 m_B) \nonumber\\
    K_{DC} &= m_D q_C^2\bigl[(1+r_2)m_C + r_2(m_A + m_B)\bigr] \nonumber\\
    K_{DD} &= -r_2 m_D q_C^2\bigl[(1+r_2)m_C + r_2(m_A + m_B)\bigr]
    \;.
\end{align}
In general, all entries $K_{ij}$ are non-zero.
However, if we choose the mobilities such that 
\begin{align}
    m_A &= m_B r_1 & m_C &= m_D r_2 \;,
\end{align}
the off-diagonal blocks coupling $(A,B)$ to $(C,D)$ vanish.
The reaction matrix then becomes block-diagonal: The first block corresponds to the reaction between $A$ and $B$, whereas the second block corresponds to the reaction between $C$ and $D$.
The reaction matrix reads
\begin{equation}
    K=D_0
\begin{pmatrix} -m_A q_A^2 &
r_1m_Aq_A^2 & 0 & 0 \\
m_A q_A^2 & -r_1m_Aq_A^2 & 0 & 0 \\ 
0 & 0 & -m_C q_C^2 &
r_2 m_C q_C^2 \\ 0 & 0 &    m_C q_C^2 & -r_2 m_C q_C^2 \end{pmatrix}
\;.
\end{equation} 

\subsection{Generalization}
This construction extends directly to $2n$ species comprising $n$ independent reaction pairs.
With the choice $m_{2s+1} = r_s m_{2s+2}$ for each pair $s = 1,\dots,n$, the reaction matrix $K_{ij}$ is block-diagonal with $n$ $2\times2$ blocks.
Each block has the form
\begin{equation}
    K^{(s)}=D_0 m_{2s+1} q_{2s+1}^2\begin{pmatrix} -1 & r_s \\ 1 & -r_s \end{pmatrix},
\end{equation}
where the component $2s+1$ carries the positive charge of flavor $s$.
Thus, an appropriate choice of the bare mobilities $m_i$ guarantees that only pairwise reactions occur within each flavor sector.
In the main text, we adopt the simplified assignment $m_A = m_C = m_E = 1$ for the positively charged components.

\section{Additional figures}
\label{sec:addition}

\subsection{Components involved in multiple reactions}
\begin{figure*}
    \centering
    \includegraphics[width=\linewidth]{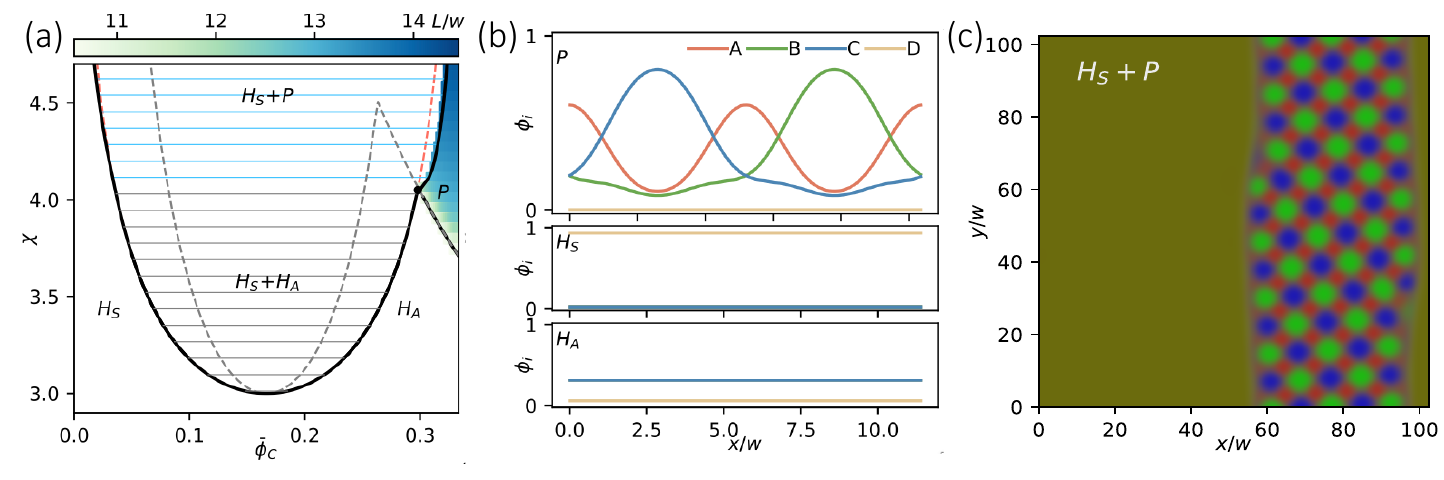}    
    \caption{(a) Phase diagram for four-component system with reactions $A \react B$ and $A \react C$.
    (b) Associated profiles of the three phases in 1D for $\phiCbar=0.333, \chi=3.98$; $\phiCbar=0.021, \chi=3.7$; and $\phiCbar=0.313, \chi=3.7$ (top to bottom).
    (c) A snapshot of the coexistence of $H_S$ and $P$ in 2D at $\phiCbar=0.15, \chi=4.1$, which shows the square lattice structure of the patterned phase.}
    \label{fig:square_lattice}
\end{figure*}

If a component is involved in more than one reaction, phases with more complicated patterns can form, but the generalized Gibbs phase rule still applies.
For example, for a system system comprising four components with reaction $A \react B$ and $A \react C$, still at most two phases can form.
If we set $k_{AB}=k_{BA}=k_{AC}=k_{CA}=0.354 k_0$, the phase diagram for one-dimensional systems is shown in \figref{fig:square_lattice}(a).
Here we denote the $D$-riched homogeneous phase $H_S$, the homogeneous phase enriched in A, B, and C is named $H_A$, and the patterned phase is denoted as $P$.
\figref{fig:square_lattice}(b) shows typical profiles for the three phases in one-dimensional systems.
In two-dimensional systems, the patterned phase can be a square lattice structure, due to the constraints $\bar{\phi}_A=\bar{\phi}_B=\bar{\phi}_C$ induced by reactions. We show the coexistence of $H_S$ and $P$ in two-dimensional systems in \figref{fig:square_lattice}(c), which shows the square lattice structure of the patterned phase.

\subsection{Amplitude of patterns in Fig.~2b}
\begin{figure}
    \centering
    \includegraphics[width=0.48\textwidth]{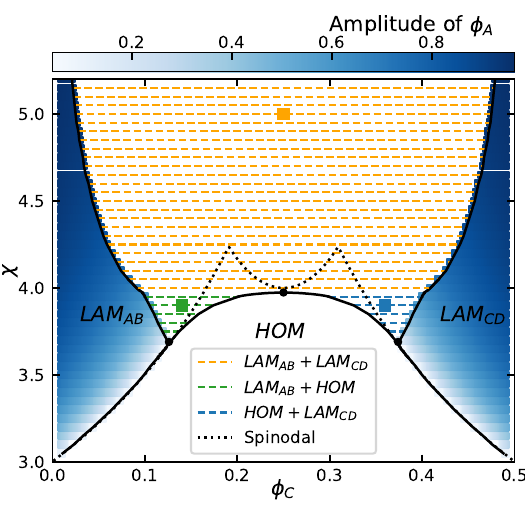}    
    \caption{
    Phase diagram associated with Fig.~2(b) in the main text, now displaying the amplitude of $\phi_A$ in the patterned phase as a color code.}
    \label{fig:amplitude_chi}
\end{figure}

Associated with the phase diagram shown in Fig.~2b of the main text, we in \figref{fig:amplitude_chi} show the amplitude of the volume fraction $\phi_A$ ($\phi_C$) in the patterned phase $LAM_{AB}$ ($LAM_{CD}$) as a function of $\chi$ and $\phi_C$.
The fact that the amplitude goes to zero at the spinodal line indicates that there is a continuous transition from homogeneous phase to patterned phase.
The amplitude increases as $\chi$ increases, indicating the patterns become more pronounced at higher $\chi$.

\subsection{Coexistence of a patterned phase and homogeneous for three components, one reaction $N{=}3$, $S{=}1$}
\begin{figure}
    \centering
    \includegraphics[width=0.48\textwidth]{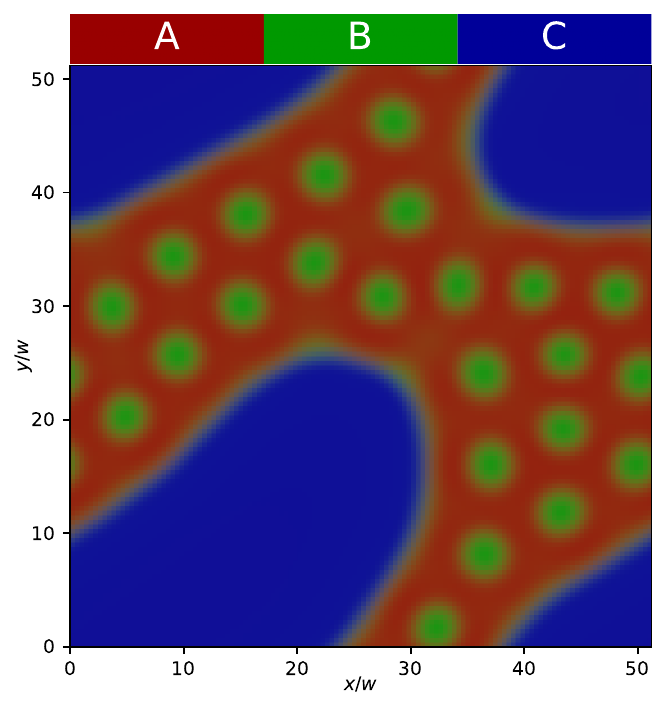}    
    \caption{A snapshot of 3-component system with one reaction at $t=1600w^2/D_0$. $\chi=5$, $r_1=3$, $k_{AB}=r_1 k_{BA}=0.38k_0$, and $m_A=r_1m_B=m_C=1$.}
    \label{fig:snapshot_3comp}
\end{figure}

For three species A, B, and C with one reaction $A \react B$, it is easy to imagine the coexistence of two homogeneous phases: one enriched in A and B, and the other enriched in C. Here we show the coexistence of a patterned AB-rich phase and a C-rich homogeneous phase in \figref{fig:snapshot_3comp}.

\subsection{Patterns in a system with seven components}
We here show additional figures for some simulations of systems with seven components.
To improve numerical stability, we add a solvent component $S$ with zero interaction with other components and zero charge to the $6$-component system mentioned in the main text.
The solvent component is passive and does not participate in reactions. Thus, the solvent component will not affect the phase behavior of the system.

\begin{figure}
    \centering
    \includegraphics[width=0.48\textwidth]{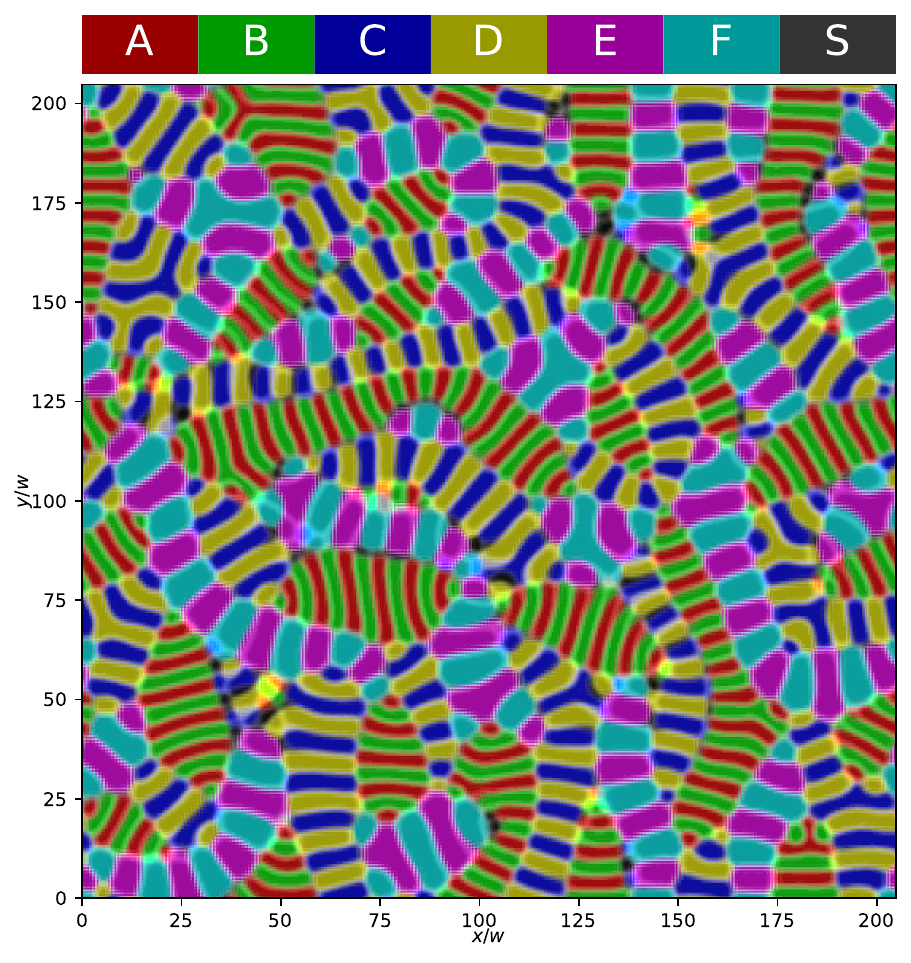}    
    \caption{A snapshot of 7-component system with three reactions at $t=1000w^2/D_0$. Reaction ratios are $r_1=r_2=r_3=1$. }
    \label{fig:snapshot_7comp_stripe}
\end{figure}

In contrast to Fig.~1 in the main text, where we used $r_1=r_2=3$ to allow hexagonal patterns to form, \figref{fig:snapshot_7comp_stripe} shows stripe patterns due to our choice of $r_1=r_2=r_3=1$. %

\begin{figure}
    \centering
    \includegraphics[width=0.48\textwidth]{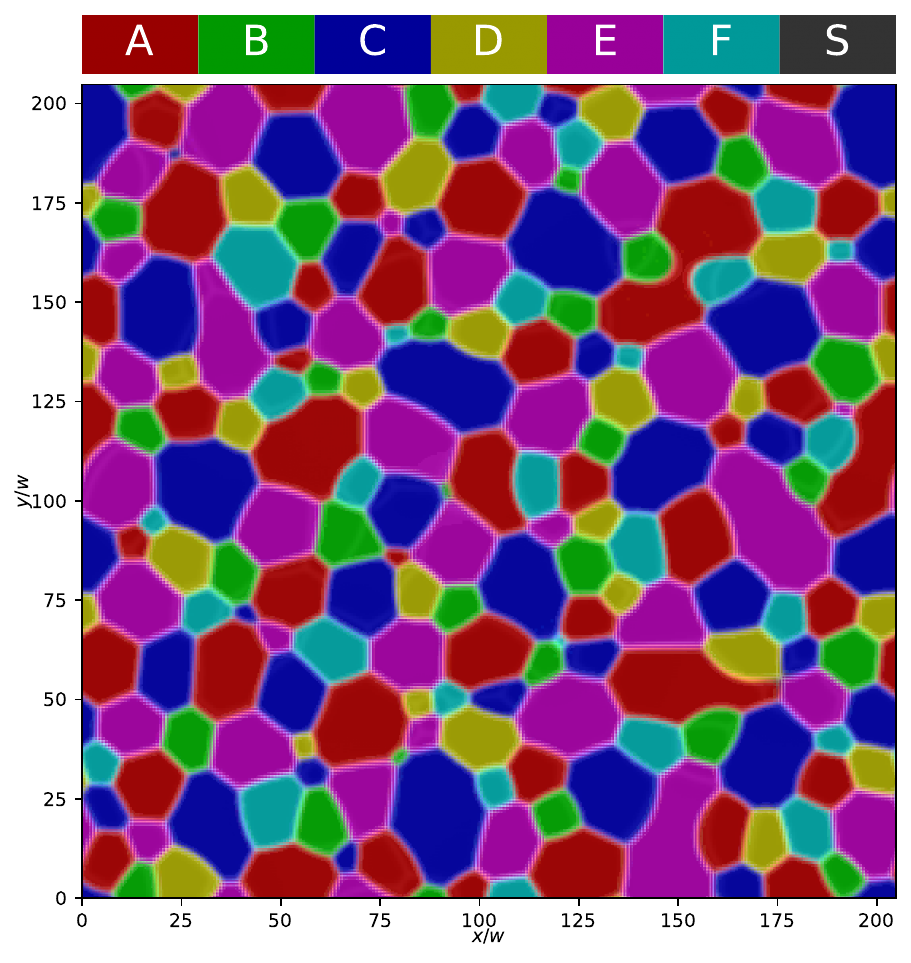}    
    \caption{A snapshot of 7-component system without reactions at $t=750w^2/D_0$.}
    \label{fig:snapshot_7comp_nocharge}
\end{figure}

\figref{fig:snapshot_7comp_nocharge} demonstrates that without reactions there is no suppression of coarsening.
Ostwald ripening occurs and the system will finally reach coexistence of 6 macrophases.

\begin{figure}
    \centering
    \includegraphics[width=0.48\textwidth]{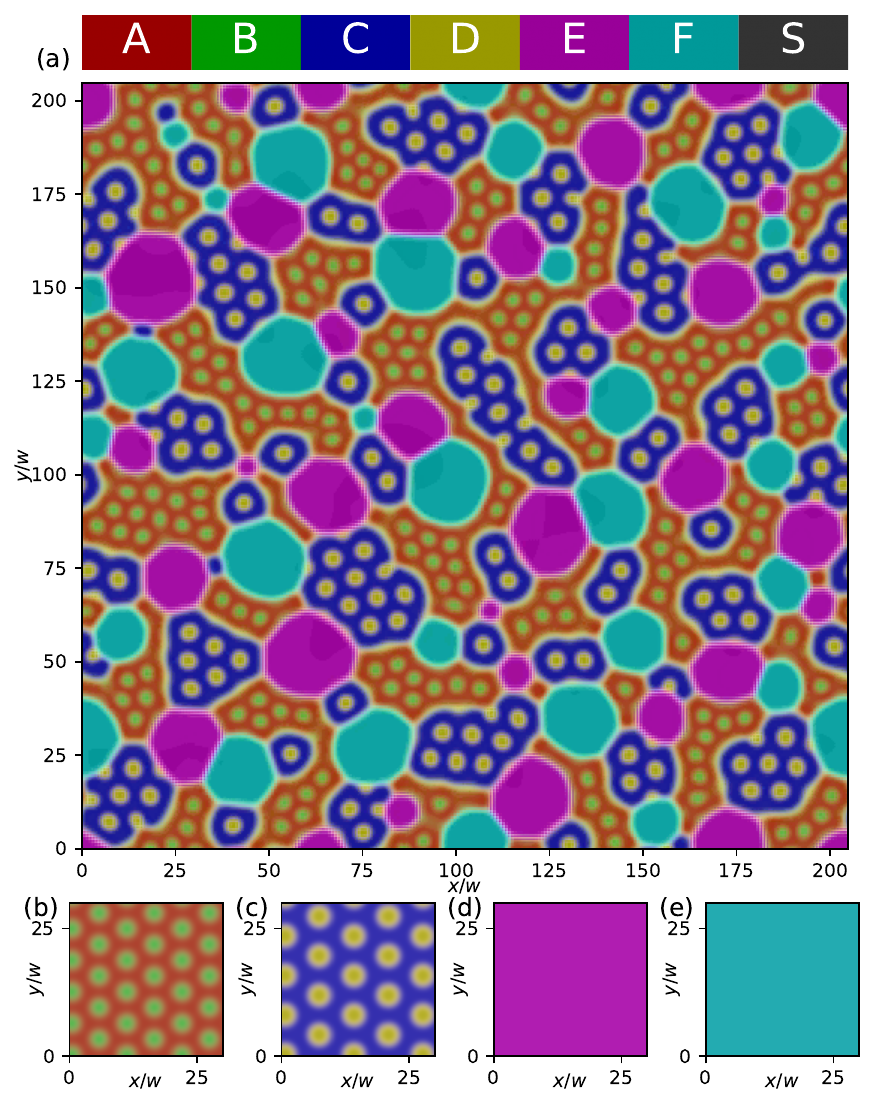}    
    \caption{A snapshot of 7-component system with only two reactions at $t=1000w^2/D_0$.}
    \label{fig:snapshot_7comp_2reactions}
\end{figure}

We also demonstrate the coexistence of four phases to test the generalized Gibbs phase rule.
When only use two reactions, $A \react B$ and $C \react D$, with $r_1=3$ and $r_2=3$, the 7-component system only has two flavors of effective charges, which can lead to two hexagonal patterned phases and two homogeneous phases.
\figref{fig:snapshot_7comp_2reactions} demonstrates the associated coexistence of two hexagonal patterned phases and two homogeneous phases.

\end{document}